\documentclass[a4paper,11pt]{article}
\usepackage{jheppub}
\usepackage{amsmath,amssymb,amsfonts}
\usepackage{subcaption}
\usepackage{booktabs}
\usepackage{placeins}
\usepackage[utf8]{inputenc}
\usepackage{lmodern}
\usepackage[T1]{fontenc}
\usepackage{graphicx} 
\usepackage{orcidlink}
\newcommand{\beq}{\begin{equation}\begin{aligned}}
\newcommand{\eeq}{\end{aligned}\end{equation}}

\usepackage{tabularx}
\usepackage{array}

\newcommand{\nocontentsline}[3]{}
\let\origcontentsline\addcontentsline
\newcommand\stoptoc{\let\addcontentsline\nocontentsline}
\newcommand\resumetoc{\let\addcontentsline\origcontentsline}

\title{Quasinormal modes response to thermodynamic phase transitions in the charged AdS black hole surrounded by perfect fluid dark matter}
\author{Long-Xiang Li,}
\emailAdd{LLXphy@163.com}
\author{Zhong-Wen Feng}
\emailAdd{zwfengphy@cwnu.edu.cn}
\affiliation{School of Physics and Astronomy, China West Normal University, Nanchong 637009, China}
\date{\today}

\abstract{We investigate how thermodynamic phase transitions are reflected in the quasinormal modes (QNMs) of charged anti-de Sitter (AdS) black holes surrounded by perfect fluid dark matter (PFDM). In the extended phase space, increasing the positive PFDM parameter raises the critical temperature and pressure while reducing the critical horizon radius. We compute the fundamental QNMs of a massless scalar perturbation using a Chebyshev pseudospectral method and analyze their evolution along isobaric and isothermal processes below the critical point. The small and large black hole branches trace clearly separated QNM trajectories and display sharply different slopes near the first-order transition, providing a dynamical signature of the branch change. Along isotherms, this evolution results from the competing effects of the horizon radius and pressure, or equivalently the AdS radius, rather than from the horizon radius alone. At the critical point, however, the QNM frequencies vary smoothly with the horizon radius and show no sharp signature of the second-order transition. Along the coexistence curve, the separation between the small and large black hole QNMs decreases as the Gibbs free energy swallowtail shrinks and vanishes at criticality. These results show that PFDM shifts both the thermodynamic phase structure and the associated QNM response, while the fundamental scalar QNM spectrum remains sensitive to the first-order small/large black hole transition.}
\makeatletter

\begin{document}
\maketitle

\section{Introduction}
The discovery that black holes have entropy and Hawking temperature revealed a deep connection between gravity, quantum theory, and statistical physics, and showed that black holes can be treated as thermodynamic systems~\cite{PhysRevD.7.2333,Hawking:1975vcx}. Once black holes are viewed in this way, their phase behavior becomes an important issue. The anti-de Sitter (AdS) spacetime provides a useful background for studying this problem because black holes can be in thermal equilibrium with the surrounding spacetime. The Hawking-Page transition first showed that an AdS black hole can undergo a phase transition with thermal AdS space~\cite{Hawking:1982dh}. For charged AdS black holes, a small/large black hole phase transition similar to the liquid-gas transition of a Van der Waals fluid was later found~\cite{Chamblin:1999tk}. In the extended phase space, the cosmological constant is treated as the thermodynamic pressure, that is $P =  - {\Lambda  \mathord{\left/ {\vphantom {\Lambda  {8\pi }}} \right. \kern-\nulldelimiterspace} {8\pi }} = {3 \mathord{\left/ {\vphantom {3 {8\pi {L^2}}}} \right. \kern-\nulldelimiterspace} {8\pi {L^2}}}$, and charged AdS black holes have an equation of state, a critical point, and a coexistence curve similar to those of a Van der Waals fluid~\cite{Kubiznak:2012wp}. These studies have been extended to many AdS black hole backgrounds, including charged, rotating, nonlinear electromagnetic, higher-curvature, and higher-dimensional black holes~\cite{Gunasekaran:2012dq,Hendi:2012um,Altamirano:2013ane,Cai:2013qga,Banerjee:2012zm,Zou:2013owa,Kubiznak:2016qmn,Wu:2023sue,Shukla:2024tkw,Promsiri:2024hrl}.

Quasinormal modes (QNMs) describe the characteristic response of black holes to linear perturbations, and their complex frequencies, $\omega=\omega_R+i\omega_I$, determine the oscillation frequency and damping rate of the perturbation~\cite{Chandrasekhar:1975zza,Berti:2009kk,Konoplya:2011qq}. Since the perturbation equation and the boundary conditions are fixed by the background geometry, the QNM spectrum encodes information about the black hole spacetime and its surrounding environment. QNMs have therefore been widely used to study black hole stability, AdS perturbation dynamics, and gravitational-wave ringdown signals~\cite{Horowitz:1999jd,LIGOScientific:2016aoc,Berti:2016lat}. Recent studies of overtones, nonlinear ringdown, and pseudospectral stability further show that QNM spectra can be sensitive to changes in the effective potential and to details of the ringdown dynamics~\cite{Giesler:2019uxc,Cotesta:2022pci,Yi:2024elj,Berti:2022xfj,Jaramillo:2020tuu,Jaramillo:2021tmt,Are_n_2023,Rosato:2024arw}. This sensitivity makes QNMs useful dynamical probes of black hole parameters and environmental effects~\cite{Cardoso:2016rao,Isi:2019aib}.

Based on the above discussion, it is natural to ask whether QNMs can be used to diagnose thermodynamic phase transitions of black holes. Previous studies have shown that QNMs can indeed carry phase transition information in several AdS black hole systems~\cite{Wang:2000gsa,Shen:2007xk,Myung:2008ze,Koutsoumbas:2008pw,Mahapatra:2016dae}. For four-dimensional RN-AdS black holes, scalar QNM frequencies exhibit a characteristic change at the first-order small/large black hole phase transition point, whereas their behavior remains smooth at the second-order critical point~\cite{Liu:2014gvf}. Similar results have been reported for higher-dimensional charged AdS black holes~\cite{Chabab:2016cem}, massive gravity~\cite{Zou:2017juz}, and regular AdS black holes~\cite{Guo:2024jhg}. In particular, the QNM evolution along the coexistence curve provides a useful perspective complementary to the usual isobaric and isothermal processes~\cite{Guo:2024jhg}. Related recent studies have also explored QNM signals, Lyapunov exponents, and optical features as dynamical or observational probes of black hole phase transitions~\cite{Hou:2025bli,Shukla:2024tkw,Promsiri:2024hrl}.

Most existing studies of the QNM-phase transition relation focus on black holes without surrounding matter fields. However, realistic black holes may be embedded in environmental matter distributions, including dark matter, which can modify both the spacetime geometry and the perturbation potential. Since the microscopic nature of dark matter is still unknown, effective descriptions are useful for exploring its gravitational influence. Perfect fluid dark matter (PFDM) provides such a phenomenological model, in which the dark matter environment is encoded in the black hole metric through an additional parameter~\cite{Rahaman:2010xs,Kiselev:2003ah}. Charged AdS black holes surrounded by PFDM have been constructed, and their thermodynamic properties have been analyzed~\cite{Xu:2016ylr,Zhang:2024jlp}. Further studies have considered rotating extensions, entropy corrections, Joule-Thomson expansion, weak cosmic censorship, and QNMs in PFDM backgrounds~\cite{Xu:2017bpz,Abbas:2023pug,Cao:2021dcq,Shaymatov:2020wtj,Ali:2025ooh,Tan:2025usr}. Despite these developments, the QNM response to thermodynamic phase transitions in a PFDM environment has not been systematically examined.

These considerations motivate us to examine whether the QNM signature of black hole thermodynamic phase transitions persists when the black hole is immersed in a PFDM environment. This question is nontrivial because the PFDM parameter affects both sides of the problem. On the thermodynamic side, it changes the equation of state, the critical quantities, and the Gibbs free energy. On the dynamical side, it modifies the background metric and hence the effective potential that governs scalar perturbations. Therefore, the PFDM background may change not only the location of the phase transition, but also the way in which this transition is reflected in the QNM spectrum.

In this work, we address this issue for charged AdS black holes surrounded by PFDM. We first review the extended phase space thermodynamics and analyze how the PFDM parameter affects the critical point and the Gibbs free energy. We then compute the fundamental massless scalar QNMs using the Chebyshev pseudospectral method. Below the critical point, we study the QNM behavior in both isobaric and isothermal processes and examine its relation to the first-order small/large black hole phase transition. For the isothermal process, we further separate the effects of the horizon radius and the pressure, or equivalently the AdS radius, in order to clarify the origin of the QNM evolution. We also investigate the QNM spectra at the critical point and along the coexistence curve, and compare the dynamical QNM response with the thermodynamic phase structure.

The paper is organized as follows. In section~\ref{SECTII}, we review the charged AdS black hole surrounded by PFDM and its thermodynamic properties in the extended phase space. In section~\ref{SECTIII}, we calculate the QNMs below the critical point and study their behavior in isobaric and isothermal processes. In section~\ref{SECTIV}, we analyze the QNM behavior at the critical point. In section~\ref{SECTV}, we investigate the QNM evolution along the coexistence curve and compare it with the Gibbs free energy structure. Finally, the conclusions and discussion are presented in section~\ref{SECTVI}.

\section{Thermodynamics and phase transitions of charged AdS black holes in perfect fluid dark matter}
\label{SECTII}
To begin with, we briefly review the charged AdS black hole surrounded by PFDM and its thermodynamic properties in the extended phase space, including the equation of state, critical point, and Gibbs free energy.

Since dark matter may form an ambient distribution around black holes, it is natural to ask how such an environment modifies black hole geometry and thermodynamics. Due to the lack of a confirmed microscopic model of dark matter, an effective fluid description is often adopted. In the PFDM model, the dark matter component is treated as a perfect fluid source, whose effect is encoded in the spacetime metric through an additional parameter. This provides a tractable way to investigate the influence of dark matter on black hole phase transitions. Following Refs.~\cite{Kiselev:2002dx,Kiselev:2003ah,Kiselev:2004vy,Li:2012zx}, the action for Einstein-Maxwell gravity coupled to perfect fluid dark matter can be written as
\begin{align}
\label{eq2.1}
S = \int {\rm d}^4x \sqrt{-g} \left( \frac{R}{16\pi G} - \frac{\Lambda}{8\pi G} + \frac{1}{4}F_{\mu\nu}F^{\mu\nu} + \mathcal{L}_{\text{DM}} \right),
\end{align}
where $G$ is Newton's gravitational constant, $\Lambda$ is the cosmological constant, $F_{\mu\nu}$ is the electromagnetic field tensor, and $\mathcal{L}_{\text{DM}}$ denotes the Lagrangian density of the dark matter fluid. For $\Lambda=-3/L^2$, the corresponding static and spherically symmetric charged black hole solution surrounded by PFDM is given by~\cite{Xu:2016ylr}
\begin{align}
\label{eq2.2}
{\rm d}s^{2} =& -f\left(r\right){\rm d}t^{2}+f\left(r\right)^{-1}{\rm d}r^{2}+r^{2}({\rm d}\theta^{2}+\sin^{2}\theta {\rm d}\phi^{2}),
\end{align}
with
\begin{align}
\label{eq2.3}
f\left(r\right) = 1 - \frac{2M}{r} + \frac{Q^2}{r^2} + \frac{r^2}{L^2} + \frac{\alpha}{r} \log\left(\frac{r}{|\alpha|}\right),
\end{align}
where $M$ and $Q$ denote the mass and electric charge of the black hole, $L$ is the AdS curvature radius, and $\alpha$ characterizes the intensity of the surrounding perfect fluid dark matter. The logarithmic term represents the correction induced by the PFDM background. When $\alpha\to0$, the metric reduces to the standard RN-AdS black hole. In the asymptotically flat limit $L\to\infty$, it further reduces to the RN black hole, and with the additional condition $Q=0$, one recovers the Schwarzschild black hole.

In the extended phase space, the cosmological constant is interpreted as a thermodynamic pressure, that is $P = {3 \mathord{\left/
 {\vphantom {3 {8\pi {L^2}}}} \right. \kern-\nulldelimiterspace} {8\pi {L^2}}}$, and the metric function~(\ref{eq2.3}) can be rewritten as
\begin{align}
\label{eq2.4}
f\left(r\right) = 1 - \frac{2M}{r} + \frac{Q^2}{r^2} + \frac{8\pi P r^2}{3} + \frac{\alpha}{r} \log\left(\frac{r}{|\alpha|}\right).
\end{align}
It is clear that the radius of the event horizon $r_h$ is determined by the largest positive root of $f\left(r_h\right)=0$. By solving this condition, the mass of the black hole can be expressed as
\begin{align}
\label{eq2.5}
M = \frac{r_h}{2} + \frac{Q^2}{2r_h} + \frac{4\pi P r_h^3}{3} + \frac{\alpha}{2} \log\left(\frac{r_h}{|\alpha|}\right).
\end{align}
Then, the corresponding Hawking temperature on the outer event horizon becomes
\begin{align}
\label{eq2.6}
T = \frac{f'\left(r_h\right)}{4\pi} = \frac{1}{4\pi r_h} \left( 1 - \frac{Q^2}{r_h^2} + 8\pi P r_h^2 + \frac{\alpha}{r_h} \right).
\end{align}
Since the PFDM parameter appears explicitly in the mass formula, it is natural to treat $\alpha$ as an additional thermodynamic variable. According to Eqs.~(\ref{eq2.4})-(\ref{eq2.6}), the first law of the spherically symmetric charged black hole with PFDM takes the form~\cite{Xu:2016ylr}
\begin{align}
\label{eq2.7}
{\rm d}M = T {\rm d}S + V {\rm d}P + \Phi {\rm d}Q + \Psi {\rm d}\alpha,
\end{align}
with the entropy $S = \pi r_h^2$, the volume $V = {{4\pi r_h^3} \mathord{\left/ {\vphantom {{4\pi r_h^3} 3}} \right. \kern-\nulldelimiterspace} 3}$, the electric potential $\Phi = Q / r_h$, and the conjugate quantity associated with the PFDM parameter is $\Psi = {\left( {{{\partial M} \mathord{\left/ {\vphantom {{\partial M} {\partial \alpha }}} \right. \kern-\nulldelimiterspace} {\partial \alpha }}} \right)_{S,P,Q}} = \frac{1}{2}\left[ {\log \left( {{{{r_h}} \mathord{\left/ {\vphantom {{{r_h}} {|\alpha |}}} \right. \kern-\nulldelimiterspace} {|\alpha |}}} \right) - 1} \right]$. Furthermore, these thermodynamic quantities satisfy the corresponding Smarr relation $M = 2TS - 2PV + \Phi Q + \alpha \Psi$. Therefore, the PFDM background not only modifies the geometry but also enlarges the thermodynamic phase space through the additional pair $\left(\alpha,\Psi\right)$.

From Eq.~(\ref{eq2.6}), the equation of state for the charged black hole with PFDM is given by
\begin{align}
\label{eq2.8}
P = \frac{T}{2r_h} - \frac{1}{8\pi r_h^2} + \frac{Q^2}{8\pi r_h^4} - \frac{\alpha}{8\pi r_h^3},
\end{align}
which has the same qualitative structure as the equation of state of a Van der Waals fluid. The electric charge provides a repulsive contribution at small horizon radius, while the PFDM correction introduces an additional term controlled by $\alpha$. This implies that the location of the critical point and the coexistence region are shifted compared with the standard RN-AdS case. To determine the critical behavior, one needs to impose the following conditions
\begin{align}
\label{eq2.9}
\left(\frac{\partial P}{\partial r_h}\right)_T = 0, \quad \left(\frac{\partial^2 P}{\partial r_h^2}\right)_T = 0.
\end{align}
According to Eqs.~(\ref{eq2.8}) and (\ref{eq2.9}), the critical horizon radius, critical temperature, and critical pressure can be obtained as
\begin{align}
r_c &= \frac{1}{2} \left( \sqrt{9\alpha^2 + 24Q^2} - 3\alpha \right), \\
T_c &= \frac{9\alpha^2 - 3\sqrt{9\alpha^2 + 24Q^2}\,\alpha + 16Q^2}{\pi \left(\sqrt{9\alpha^2 + 24Q^2} - 3\alpha\right)^3}, \\
P_c &= \frac{3\alpha^2 - \sqrt{9\alpha^2 + 24Q^2}\,\alpha + 6Q^2}{\pi \left(\sqrt{9\alpha^2 + 24Q^2} - 3\alpha\right)^4}.
\end{align}
In the limit $\alpha\to0$, these quantities reduce to the classical critical values of the RN-AdS black hole $r_c=\sqrt{6}Q$, ${T_c} = {{\sqrt 6 } \mathord{\left/ {\vphantom {{\sqrt 6 } {18\pi Q}}} \right. \kern-\nulldelimiterspace} {18\pi Q}}$, and ${P_c} = {1 \mathord{\left/
 {\vphantom {1 {96\pi {Q^2}}}} \right. \kern-\nulldelimiterspace} {96\pi {Q^2}}}$. Thus, the PFDM parameter continuously deforms the standard RN-AdS phase structure. For the positive PFDM parameters considered in this work, increasing $\alpha$ decreases the critical radius, while increasing both the critical temperature and the critical pressure. This means that the PFDM environment changes the thermodynamic scale at which the small/large black hole transition occurs.

To determine the globally preferred phase, we further consider the Gibbs free energy. For the charged black hole with PFDM, the Gibbs free energy is obtained from $G=M-TS$ as~\cite{Xu:2016ylr,Ali:2025ooh}
\begin{align}
\label{eq2.13}
G = M - TS = \frac{r_h}{4} - \frac{2\pi}{3} P r_h^3 + \frac{3Q^2}{4r_h} + \frac{\alpha}{2} \log \left( \frac{r_h}{|\alpha|} \right) - \frac{\alpha}{4}.
\end{align}

In the following numerical analysis, we focus on positive values of the PFDM parameter, which correspond to the positive energy density branch of the surrounding fluid and are most relevant to the dark matter interpretation~\cite{Kiselev:2003ah,Rahaman:2010xs}. The negative-$\alpha$ regime may have different horizon and energy-condition properties and will not be considered here~\cite{Xu:2016ylr}. We fix $Q=0.75$ and choose $\alpha=0.1,0.3,0.5$ as representative positive values to study the effect of PFDM  on the phase transition and the QNM response~\cite{Ali:2025ooh}.

\section{QNM behavior below the critical point}
\label{SECTIII}
\subsection{Scalar perturbations and QNMs}
\label{SECT3.1}
We now consider a massless scalar perturbation in the charged AdS black hole surrounded by PFDM. The perturbation is treated in the test field approximation, so its backreaction on the spacetime, electromagnetic field, and PFDM background is neglected. The scalar field satisfies the Klein-Gordon equation
\begin{align}
\label{eq3.1}
\frac{1}{\sqrt{-g}} \partial_\mu \left( \sqrt{-g} g^{\mu\nu} \partial_\nu \varphi \right) = 0,
\end{align}
where $\varphi(t,r,\theta,\phi)$ denotes the massless scalar perturbation. In the spherically symmetric spacetime, the scalar field can be decomposed as
\begin{align}
\label{eq3.2}
\varphi(t,r,\theta,\phi) = \sum_{lm} \frac{\psi_{lm}(r)}{r} e^{-i\omega t} Y_{lm}(\theta,\phi),
\end{align}
where $Y_{lm}(\theta,\phi)$ are spherical harmonics, $\omega$ is the complex QNM frequency, and $\psi_{lm}(r)$ is the radial wave function. By introducing the tortoise coordinate ${\rm d}r_*={\rm d}r/f\left(r\right)$, the radial equation is written in the Schr\"odinger-like form
\begin{align}
\label{eq3.3}
\frac{{\rm d}^2\psi}{{\rm d}r_*^2} + \left[ \omega^2 - V\left(r\right) \right] \psi = 0,
\end{align}
with the effective potential
\begin{align}
\label{eq3.4}
V\left(r\right) = f\left(r\right) \left[ \frac{l\left(l+1\right)}{r^2} + \frac{f'\left(r\right)}{r} \right].
\end{align}
For a given set of black hole parameters $\left(r_h,P,Q,\alpha \right)$, the mass parameter $M$ in $f\left(r\right)$ is fixed by the horizon condition $f\left(r_h\right)=0$. Therefore, once $r_h$, $P$, $Q$, and $\alpha$ are specified, the background geometry and the effective potential are determined.

In this study, we focus on the fundamental $s$-wave mode, with the overtone number $n=0$ and the angular quantum number $l=0$. The QNM boundary conditions are imposed as purely ingoing waves at the event horizon and vanishing scalar perturbations at the AdS boundary. With the time dependence $e^{-i\omega t}$, stable modes satisfy $\operatorname{Im}(\omega)<0$. QNM frequencies can be computed by several methods, such as the asymptotic iteration method~\cite{Cho:2009cj}, the continued fraction method~\cite{Leaver:1985ax}, and the Horowitz-Hubeny method~\cite{Horowitz:1999jd}. In this work, we employ the Chebyshev pseudospectral method to obtain the frequencies numerically. To implement this method, we first factor out the ingoing behavior at the event horizon by writing
\begin{align}
\label{eq3.5}
\psi(r)=e^{-i\omega r_*}\phi(r).
\end{align}
The radial equation then becomes
\begin{align}
\label{eq3.6}
f(r)\phi''\left(r\right)+\left[f'(r)-2i\omega\right]\phi'\left(r\right)
-\left[\frac{l\left(l+1\right)}{r^2}+\frac{f'\left(r\right)}{r}\right]\phi\left(r\right)=0.
\end{align}
To solve Eq.~(\ref{eq3.6}) numerically by the Chebyshev pseudospectral method, we first map the radial domain $r\in[r_h,\infty)$ to a finite interval. This is done by introducing the compact coordinate $z=r_h/r$, so that the AdS boundary and the event horizon are located at $z=0$ and $z=1$, respectively. Under this transformation, the radial derivatives become
${\rm d}/{{\rm d}r}=- \left(z^2/r_h\right)\left ({\rm d}/{{\rm d}z}\right )$ and ${\rm d}^2/{\rm d}r^2=\left(z^4/{r^2_h}\right)\left({\rm d^2}/{{\rm d}z^2}\right)+\left(2 z^3/{r^2_h}\right) \left({\rm d}/{{\rm d}z}\right)$. The interval $z\in[0,1]$ is then discretized by Chebyshev collocation points $z_j=\left[1+\cos\left(j\pi/N\right)\right]/2$, with $j=0,1,\ldots,N$. At these collocation points, the derivative operators are replaced by the Chebyshev differentiation matrices $D$ and $D^2$, and Eq.~(\ref{eq3.6}) is converted into a matrix eigenvalue problem of the form 
\begin{align}
\label{eq3.7}
\left(M_0+\omega M_1\right)\boldsymbol{\phi}=0,
\end{align}
where $\boldsymbol{\phi}=(\phi_0,\phi_1,\ldots,\phi_N)^T$ denotes the values of $\phi$ at the collocation points, with $\phi_j=\phi\left(z_j\right)$. The matrix $M_0$ contains the background functions and the $\omega$-independent derivative terms, while $M_1$ collects the terms linear in $\omega$. The boundary conditions are imposed by replacing the corresponding boundary rows. At the AdS boundary, $z=0$, we impose $\phi\left(z=0\right)=0$. At the event horizon, $z=1$, the ingoing behavior has already been factored out in Eq.~(\ref{eq3.5}), and the remaining function $\phi$ is required to be regular. The horizon boundary row is obtained from the $r\to r_h$ limit of Eq.~(\ref{eq3.6}). The allowed QNM frequencies are then obtained from $M_0\boldsymbol{\phi}=-\omega  M_1\boldsymbol{\phi}$. We use $N=120$, corresponding to $121$ Chebyshev collocation points. Convergence tests with $60\leq N\leq150$ show that the results at $N=120$ agree with those at $N=150$ within $10^{-6}$ for the fundamental mode.

\subsection{QNM behavior in the isobaric process}

\begin{figure}[htbp]
\centering
\begin{subfigure}[t]{0.49\textwidth}
\centering
\includegraphics[width=\textwidth]{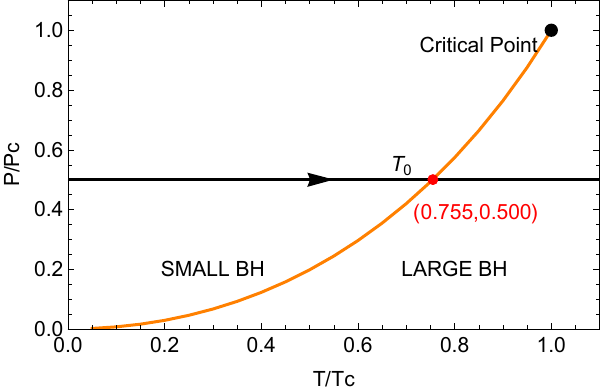}
\caption{$\alpha=0.1$}
\label{fig:a}
\end{subfigure}
\hfill
\begin{subfigure}[t]{0.49\textwidth}
\centering
\includegraphics[width=\textwidth]{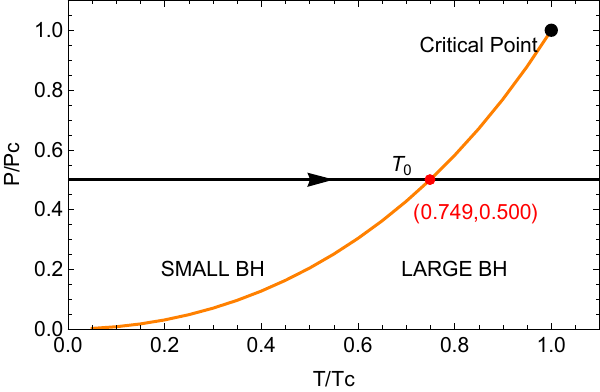}
\caption{$\alpha=0.3$}
\label{fig:b}
\end{subfigure}
\vspace{1em}
\begin{subfigure}[t]{0.49\textwidth}
\centering
\includegraphics[width=\textwidth]{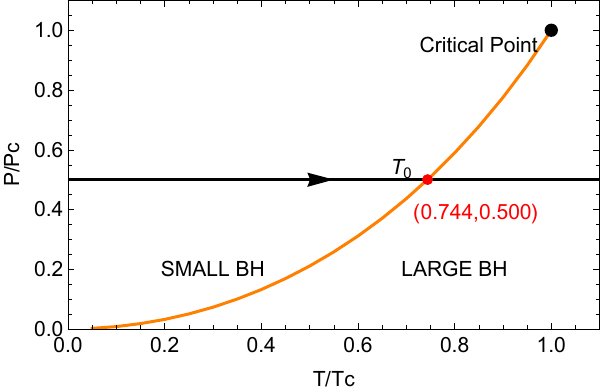}
\caption{$\alpha=0.5$}
\label{fig:c}
\end{subfigure}
\caption{The coexistence curve in the $(P,T)$ plane together with the isobar $P/P_c=0.5$. The intersection determines the first-order small/large black hole transition temperature $T_0$. (a) $\alpha=0.1$, (b) $\alpha=0.3$, (c) $\alpha=0.5$.} 
\label{fig:1}
\end{figure}
To study the QNM behavior in an isobaric process, we first determine where the chosen isobar intersects the small/large black hole coexistence line. For this purpose, Fig.~\ref{fig:1} shows the coexistence line in the $(P,T)$ plane together with the fixed reduced pressure $P/P_c=0.5$ for $\alpha=0.1$, $0.3$, and $0.5$. The coexistence line is obtained from Maxwell's equal area law. For each value of $\alpha$, the intersection, marked by a red dot, gives the first-order small/large black hole transition point. The corresponding transition temperatures are $T_0=0.0493939$, $0.0637975$, and $0.0833983$ for $\alpha=0.1$, $0.3$, and $0.5$, respectively. Thus, at fixed $P/P_c=0.5$, increasing the positive PFDM parameter $\alpha$ shifts the transition to a higher absolute temperature.

\begin{table}[htbp]
\centering
\begin{subtable}[t]{0.45\textwidth}
\centering
\begin{tabular}{ccc}
\toprule
$T$ & $r_h$ & $\omega$ \\
\midrule
0.0450 & 0.917828 & $0.412287 - 0.077283i$ \\
0.0460 & 0.930634 & $0.412202 - 0.077776i$ \\
0.0470 & 0.944662 & $0.412116 - 0.078344i$ \\
0.0480 & 0.960184 & $0.412029 - 0.079006i$ \\
0.0490 & 0.977575 & $0.411939 - 0.079792i$ \\
\midrule
0.0500 & 4.308101 & $0.446899 - 0.353870i$ \\
0.0510 & 4.594560 & $0.455082 - 0.377707i$ \\
0.0520 & 4.851115 & $0.462874 - 0.399001i$ \\
0.0530 & 5.088322 & $0.470436 - 0.418652i$ \\
0.0540 & 5.311775 & $0.477853 - 0.437134i$ \\
\bottomrule
\end{tabular}
\caption{$\alpha=0.1$}
\label{tab:a}
\end{subtable}
\hfill
\begin{subtable}[t]{0.45\textwidth}
\centering
\begin{tabular}{ccc}
\toprule
$T$ & $r_h$ & $\omega$ \\
\midrule
0.0590 & 0.803299 & $0.534145 - 0.103872i$ \\
0.0600 & 0.812029 & $0.534039 - 0.104413i$ \\
0.0610 & 0.821422 & $0.533393 - 0.105017i$ \\
0.0620 & 0.831594 & $0.533819 - 0.105717i$ \\
0.0630 & 0.842695 & $0.533703 - 0.106477i$ \\
\midrule
0.0640 & 3.471635 & $0.570990 - 0.446643i$ \\
0.0650 & 3.667782 & $0.579409 - 0.472457i$ \\
0.0660 & 3.843263 & $0.587395 - 0.495496i$ \\
0.0670 & 4.005166 & $0.595116 - 0.516712i$ \\
0.0680 & 4.157315 & $0.602663 - 0.536618i$ \\
\bottomrule
\end{tabular}
\caption{$\alpha=0.3$}
\label{tab:b}
\end{subtable}
\vspace{1em}
\begin{subtable}[t]{0.45\textwidth}
\centering
\begin{tabular}{ccc}
\toprule
$T$ & $r_h$ & $\omega$ \\
\midrule
0.0790 & 0.710165 & $0.700190 - 0.141545i$ \\
0.0800 & 0.716357 & $0.700057 - 0.142179i$ \\
0.0810 & 0.722942 & $0.699921 - 0.142874i$ \\
0.0820 & 0.729975 & $0.699780 - 0.143641i$ \\
0.0830 & 0.737528 & $0.699632 - 0.144493i$ \\
\midrule
0.0840 & 2.958898 & $0.747377 - 0.600731i$ \\
0.0850 & 3.078849 & $0.755582 - 0.625834i$ \\
0.0860 & 3.189341 & $0.763489 - 0.648915i$ \\
0.0870 & 3.292967 & $0.771194 - 0.670529i$ \\
0.0880 & 3.391330 & $0.778752 - 0.691018i$ \\
\bottomrule
\end{tabular}
\caption{$\alpha=0.5$}
\label{tab:c}
\end{subtable}
\caption{The QNM frequencies of the small and large black hole branches as functions of $T$ and $r_h$ during the isobaric process with $P/P_c = 0.5$. The entries above the horizontal line correspond to the small black hole branch, while those below it correspond to the large black hole branch. (a) $\alpha=0.1$, (b) $\alpha=0.3$, (c) $\alpha=0.5$.}
\label{table1}
\end{table}

\begin{figure}[htbp]
\centering
\begin{subfigure}[t]{\textwidth}
\centering
\begin{tabular}{cc}
\includegraphics[width=0.5\textwidth]{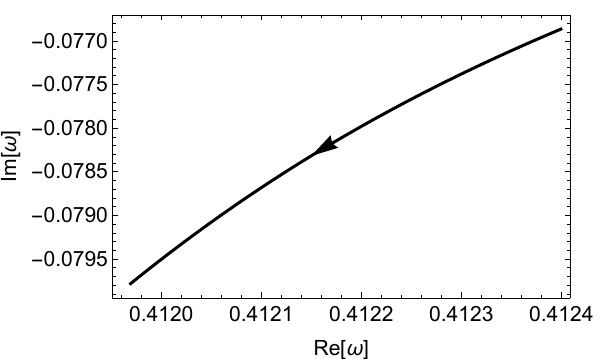} &
\includegraphics[width=0.5\textwidth]{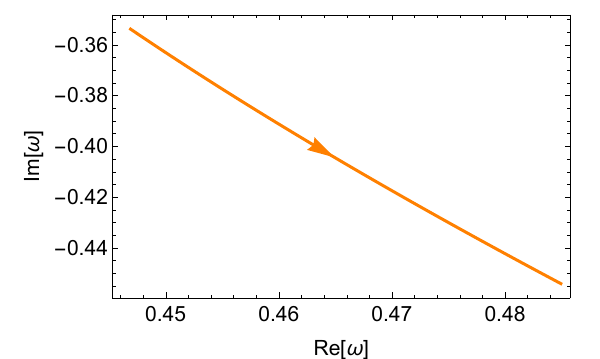}
\end{tabular}
\caption{$\alpha = 0.1$}
\label{fig:2a}
\end{subfigure}
\vspace{1em}
\begin{subfigure}[t]{\textwidth}
\centering
\begin{tabular}{cc}
\includegraphics[width=0.5\textwidth]{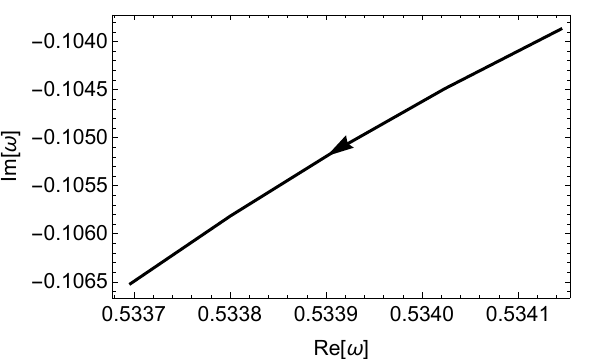} &
\includegraphics[width=0.5\textwidth]{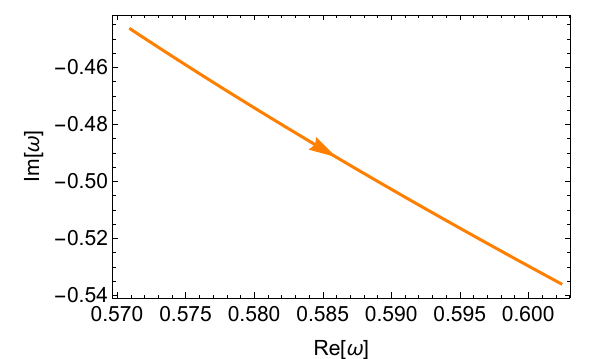}
\end{tabular}
\caption{$\alpha = 0.3$}
\label{fig:2b}
\end{subfigure}
\vspace{1em}
\begin{subfigure}[t]{1.0\textwidth}
\centering
\begin{tabular}{cc}
\includegraphics[width=0.5\textwidth]{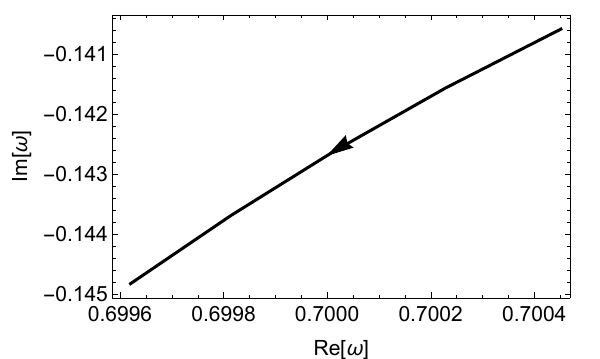} &
\includegraphics[width=0.5\textwidth]{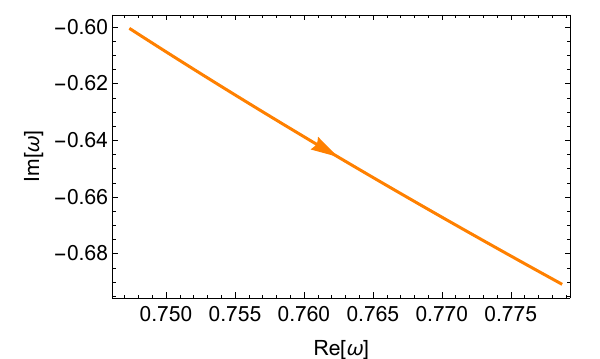}
\end{tabular}
\caption{$\alpha = 0.5$}
\label{fig:2c}
\end{subfigure}
\caption{The QNM frequencies of the small and large black hole branches as functions of the horizon radius $r_h$ during the isobaric process with $P/P_c=0.5$. In each panel, the left plot corresponds to the small black hole branch, and the right plot to the large black hole branch. The arrows indicate the direction of increasing $r_h$. (a) $\alpha=0.1$, (b) $\alpha=0.3$, (c) $\alpha=0.5$.}
\label{fig:2}
\end{figure}

After identifying the transition points, we examine the QNM frequencies on the small and large black hole branches, as listed in Table~\ref{table1}. Since the pressure is fixed, the Hawking temperature $T$ and the horizon radius $r_h$ vary simultaneously along each branch. The entries above the horizontal line correspond to the small black hole branch, while those below it correspond to the large black hole branch. On the small black hole branch, as $T$ increases toward the transition point, $r_h$ also increases. In this region, the real part of the QNM frequency changes only weakly, while $|\operatorname{Im}(\omega)|$ increases slightly. Thus, smaller black holes have weaker damping, whereas the oscillation frequency remains nearly unchanged. On the large black hole branch, increasing $T$ above the transition point also increases $r_h$, and both $\operatorname{Re}(\omega)$ and $|\operatorname{Im}(\omega)|$ increase. This indicates that larger black holes oscillate faster and damp more rapidly. This trend can be qualitatively understood from the increase in the horizon size, which enhances the absorption of scalar perturbations. Comparing different PFDM parameters in the corresponding transition regions, one also finds that larger $\alpha$ generally gives larger $\operatorname{Re}(\omega)$ and larger $|\operatorname{Im}(\omega)|$. Thus, the PFDM background affects not only the thermodynamic phase structure but also the dynamical response of the black hole to scalar perturbations.

To visualize the trends identified in Table~\ref{table1}, we plot the QNM frequencies as functions of $r_h$ in Fig.~\ref{fig:2}. The QNM curves of the small and large black hole branches are clearly separated. At the first-order transition point, the thermodynamically preferred state changes from the small black hole branch to the large black hole branch, and the corresponding QNM frequency changes discontinuously. This branch change, together with the distinct slopes of the two QNM curves, provides a dynamical signature of the first-order small/large black hole phase transition. Therefore, in the isobaric process, the QNM spectrum is sensitive to the underlying small/large black hole branch structure and provides a dynamical complement to the thermodynamic analysis.

\subsection{QNM behavior in the isothermal process}
\begin{figure}[htbp]
\centering
\begin{subfigure}[t]{0.49\textwidth}
\centering
\includegraphics[width=\textwidth]{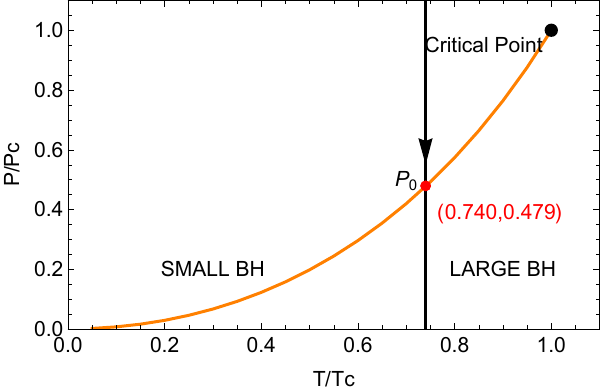}
\caption{$\alpha=0.1$}
\label{fig:3a}
\end{subfigure}
\hfill
\begin{subfigure}[t]{0.49\textwidth}
\centering
\includegraphics[width=\textwidth]{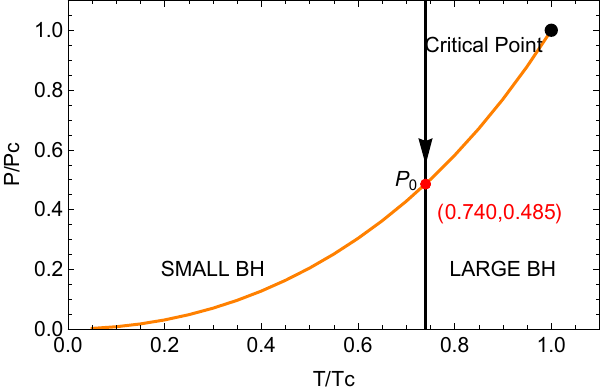}
\caption{$\alpha=0.3$}
\label{fig:3b}
\end{subfigure}
\vspace{1em}
\begin{subfigure}[t]{0.49\textwidth}
\centering
\includegraphics[width=\textwidth]{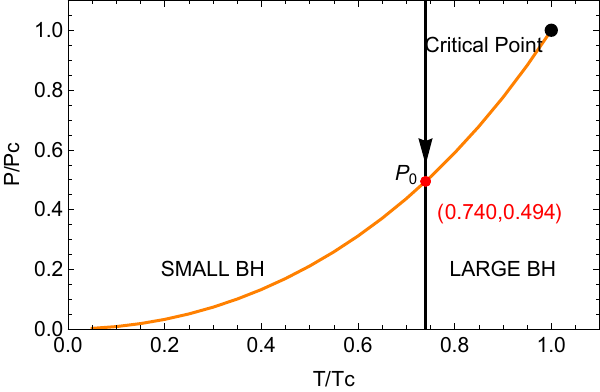}
\caption{$\alpha=0.5$}
\label{fig:3c}
\end{subfigure}
\caption{The coexistence curve in the $(P,T)$ plane together with the isotherm $T/T_c=0.74$. The intersection determines the first-order small/large black hole transition pressure $P_0$. (a) $\alpha=0.1$, (b) $\alpha=0.3$, (c) $\alpha=0.5$.}
\label{fig:3}
\end{figure}

We next turn to the isothermal process. As in the isobaric case, the first step is to locate the intersection between the chosen isotherm and the small/large black hole coexistence line. Figure~\ref{fig:3} shows the coexistence line in the $(P,T)$ plane together with the fixed reduced temperature $T/T_c=0.74$ for $\alpha=0.1$, $0.3$, and $0.5$. The coexistence line is obtained from Maxwell's equal area law. For each value of $\alpha$, the intersection, marked by a red dot, gives the first-order small/large black hole transition point. The corresponding transition pressures are $P_0=0.00351106$, $0.00561897$, and $0.00904392$ for $\alpha=0.1$, $0.3$, and $0.5$, respectively. Thus, at fixed $T/T_c=0.74$, increasing the positive PFDM parameter $\alpha$ shifts the transition to a higher pressure.

The QNM data along the isotherm are summarized in Table~\ref{table2}. The upper and lower parts of each subtable represent the small and large black hole branches, respectively. In contrast to the isobaric case, the pressure now changes along the evolution path. As $P$ decreases, $r_h$ increases on both branches. On the small black hole branch, this evolution is accompanied by a decrease in both $\operatorname{Re}(\omega)$ and $|\operatorname{Im}(\omega)|$, so the perturbation oscillates and decays more slowly. On the large black hole branch, however, $\operatorname{Re}(\omega)$ is nearly unchanged, while $|\operatorname{Im}(\omega)|$ increases noticeably, indicating a stronger damping rate. For larger $\alpha$, the representative transition region moves to smaller horizon radii, while the QNM frequencies are shifted to larger $\operatorname{Re}(\omega)$ and larger $|\operatorname{Im}(\omega)|$. These features show that the isothermal QNM behavior cannot be attributed to the horizon radius alone, and motivate the separate analysis of the effects of $r_h$ and $P$ below.

\begin{table}[htbp]
\centering
\begin{subtable}[t]{0.45\textwidth}
\centering
\begin{tabular}{ccc}
\toprule
$P$ & $r_h$ & $\omega$ \\
\midrule
0.0040 & 0.956983 & $0.427024 - 0.087964i$ \\
0.0039 & 0.960060 & $0.422482 - 0.085304i$ \\
0.0038 & 0.963224 & $0.417875 - 0.082650i$ \\
0.0037 & 0.966481 & $0.413199 - 0.080001i$ \\
0.0036 & 0.969836 & $0.408452 - 0.077357i$ \\
\midrule
0.0035 & 4.283238 & $0.432872 - 0.334881i$ \\
0.0034 & 4.577826 & $0.432564 - 0.347926i$ \\
0.0033 & 4.868587 & $0.431949 - 0.359321i$ \\
0.0032 & 5.161877 & $0.431119 - 0.369559i$ \\
0.0031 & 5.461963 & $0.430131 - 0.378932i$ \\
\bottomrule
\end{tabular}
\caption{$\alpha=0.1$}
\label{tab:2a}
\end{subtable}
\hfill
\begin{subtable}[t]{0.45\textwidth}
\centering
\begin{tabular}{ccc}
\toprule
$P$ & $r_h$ & $\omega$ \\
\midrule
0.0060 & 0.838283 & $0.542173 - 0.111266i$ \\
0.0059 & 0.840172 & $0.538256 - 0.109027i$ \\
0.0058 & 0.842100 & $0.534302 - 0.106791i$ \\
0.0057 & 0.844068 & $0.530311 - 0.104558i$ \\
0.0056 & 0.846079 & $0.526281 - 0.102329i$ \\
\midrule
0.0055 & 3.681025 & $0.561542 - 0.450521i$ \\
0.0054 & 3.830359 & $0.561125 - 0.460554i$ \\
0.0053 & 3.978982 & $0.560570 - 0.469809i$ \\
0.0052 & 4.128122 & $0.559905 - 0.478439i$ \\
0.0051 & 4.278727 & $0.559150 - 0.486551i$ \\
\bottomrule
\end{tabular}
\caption{$\alpha=0.3$}
\label{tab:2b}
\end{subtable}
\vspace{1em}
\begin{subtable}[t]{0.45\textwidth}
\centering
\begin{tabular}{ccc}
\toprule
$P$ & $r_h$ & $\omega$ \\
\midrule
0.0095 & 0.733375 & $0.710517 - 0.150628i$ \\
0.0094 & 0.734481 & $0.707204 - 0.148727i$ \\
0.0093 & 0.735603 & $0.703872 - 0.146829i$ \\
0.0092 & 0.736741 & $0.700521 - 0.144933i$ \\
0.0091 & 0.737895 & $0.697151 - 0.143039i$ \\
\midrule
0.0090 & 2.954480 & $0.737998 - 0.588234i$ \\
0.0089 & 3.032004 & $0.737752 - 0.597340i$ \\
0.0088 & 3.108583 & $0.737412 - 0.605894i$ \\
0.0087 & 3.184615 & $0.736995 - 0.613982i$ \\
0.0086 & 3.260411 & $0.736510 - 0.621671i$ \\
\bottomrule
\end{tabular}
\caption{$\alpha=0.5$}
\label{tab:2c}
\end{subtable}
\caption{The QNM frequencies of the small and large black hole branches as functions of $P$ and $r_h$ during the isothermal process with $T/T_c = 0.74$. The entries above the horizontal line correspond to the small black hole branch, while those below correspond to the large black hole branch. (a) $\alpha=0.1$, (b) $\alpha=0.3$, (c) $\alpha=0.5$.}
\label{table2}
\end{table}

\begin{figure}[htbp]
\centering
\begin{subfigure}[t]{\textwidth}
\centering
\begin{tabular}{cc}
\includegraphics[width=0.5\textwidth]{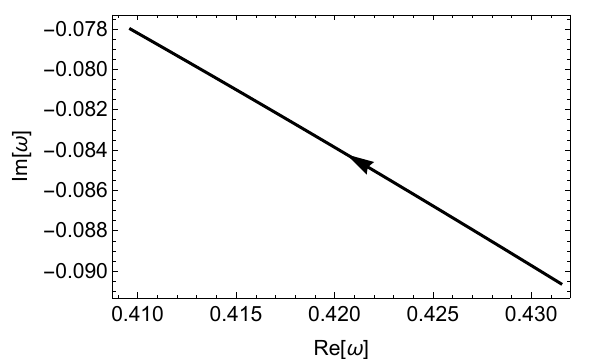} &
\includegraphics[width=0.5\textwidth]{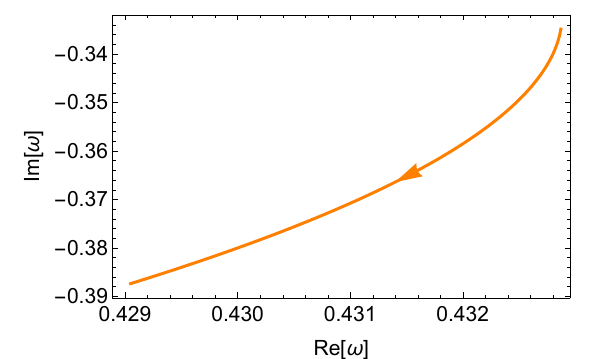}
\end{tabular}
\caption{$\alpha = 0.1$}
\label{fig:4a}
\end{subfigure}
\vspace{1em}
\begin{subfigure}[t]{\textwidth}
\centering
\begin{tabular}{cc}
\includegraphics[width=0.5\textwidth]{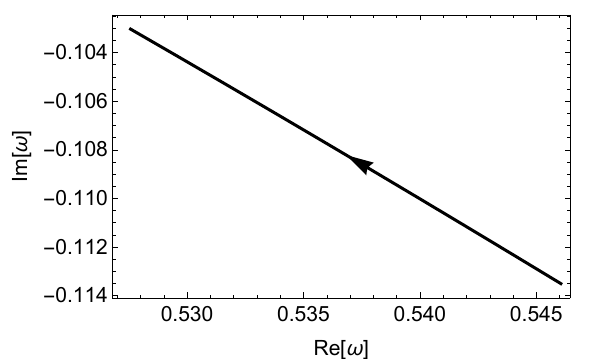} &
\includegraphics[width=0.5\textwidth]{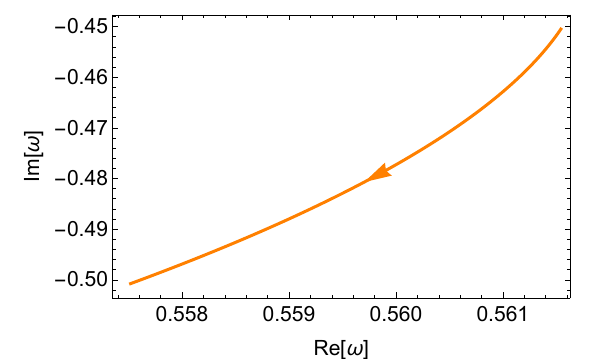}
\end{tabular}
\caption{$\alpha = 0.3$}
\label{fig:4b}
\end{subfigure}
\vspace{1em}
\begin{subfigure}[t]{\textwidth}
\centering
\begin{tabular}{cc}
\includegraphics[width=0.5\textwidth]{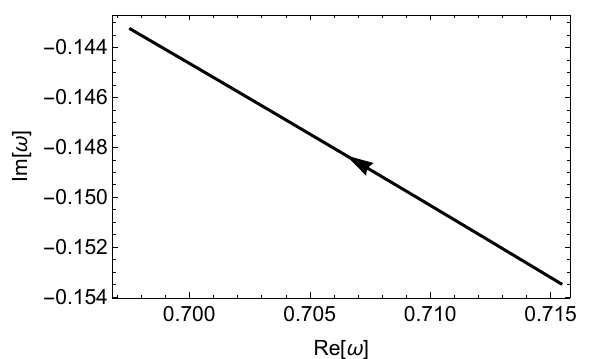} &
\includegraphics[width=0.5\textwidth]{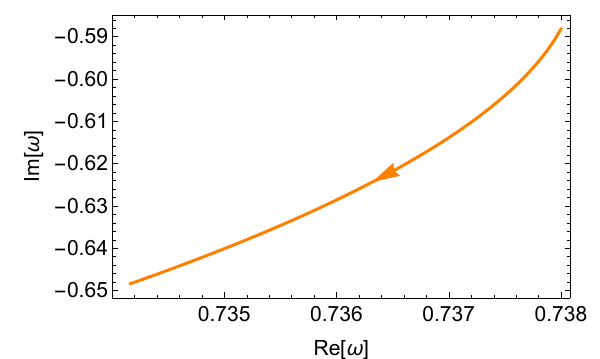}
\end{tabular}
\caption{$\alpha = 0.5$}
\label{fig:4c}
\end{subfigure}
\caption{The QNM frequencies of the small and large black hole branches as functions of the horizon radius $r_h$ during the isothermal process with $T/T_c=0.74$. In each panel, the left plot corresponds to the small black hole branch, and the right plot to the large black hole branch. The arrows indicate the direction of increasing $r_h$. (a) $\alpha=0.1$, (b) $\alpha=0.3$, (c) $\alpha=0.5$.}
\label{fig:4}
\end{figure}
The corresponding QNM spectra are displayed as functions of the horizon radius in Fig.~\ref{fig:4}. The small black hole and large black hole branches form two disjoint curves, and at the transition pressure $P_0$ the QNM frequency undergoes an abrupt jump from one branch to the other. This pronounced discontinuity, together with the markedly different slopes of the two branches, shows that the fundamental scalar QNM spectrum is strongly correlated with the first-order small/large black hole phase transition in the isothermal process.

\subsection{Competing effects of $r_h$ and $P$ on the QNM behavior}

In the isothermal process, the QNM frequencies are influenced by both the horizon radius and the pressure. Because $P$ varies along the isotherm, the AdS radius $L$, related to the pressure by $P=3/(8\pi L^2)$, varies as well. The QNM behavior shown in Table~\ref{table2} and Fig.~\ref{fig:4} therefore reflects the combined effects of $r_h$ and the AdS scale. To distinguish these contributions, we first fix $P$ and vary $r_h$, and then fix $r_h$ and vary $P$.
\begin{table}[htbp]
\centering
\begin{subtable}[t]{0.45\textwidth}
\centering
\begin{tabular}{ccc}
\toprule
$P$ & $r_h$ & $\omega$ \\
\midrule
0.0040 & 0.9370 & $0.427123 - 0.087064i$ \\
0.0040 & 0.9470 & $0.427072 - 0.087506i$ \\
0.0040 & 0.9570 & $0.427024 - 0.087965i$ \\
0.0040 & 0.9670 & $0.426981 - 0.088440i$ \\
0.0040 & 0.9770 & $0.426940 - 0.088930i$ \\
\bottomrule
\end{tabular}
\caption{$\alpha=0.1$}
\label{tab:3a}
\end{subtable}
\hfill
\begin{subtable}[t]{0.45\textwidth}
\centering
\begin{tabular}{ccc}
\toprule
$P$ & $r_h$ & $\omega$ \\
\midrule
0.0059 & 0.8200 & $0.538468 - 0.107630i$ \\
0.0059 & 0.8300 & $0.538360 - 0.108308i$ \\
0.0059 & 0.8400 & $0.538258 - 0.109014i$ \\
0.0059 & 0.8500 & $0.538160 - 0.109748i$ \\
0.0059 & 0.8600 & $0.538066 - 0.110509i$ \\
\bottomrule
\end{tabular}
\caption{$\alpha=0.3$}
\label{tab:3b}
\end{subtable}
\vspace{1em}
\begin{subtable}[t]{0.45\textwidth}
\centering
\begin{tabular}{ccc}
\toprule
$P$ & $r_h$ & $\omega$ \\
\midrule
0.0093 & 0.7150 & $0.704283 - 0.144567i$ \\
0.0093 & 0.7250 & $0.704079 - 0.145638i$ \\
0.0093 & 0.7350 & $0.703884 - 0.146760i$ \\
0.0093 & 0.7450 & $0.703696 - 0.147932i$ \\
0.0093 & 0.7550 & $0.703515 - 0.149154i$ \\
\bottomrule
\end{tabular}
\caption{$\alpha=0.5$}
\label{tab:3c}
\end{subtable}
\caption{The QNM frequencies of the small black hole branch as functions of $r_h$ at fixed pressure. (a) $\alpha=0.1$, (b) $\alpha=0.3$, (c) $\alpha=0.5$.}
\label{table3}
\end{table}

\begin{table}[htbp]
\centering
\begin{subtable}[t]{0.45\textwidth}
\centering
\begin{tabular}{ccc}
\toprule
$P$ & $r_h$ & $\omega$ \\
\midrule
0.0034 & 4.557 & $0.432023 - 0.346320i$ \\
0.0034 & 4.567 & $0.432282 - 0.347091i$ \\
0.0034 & 4.577 & $0.432542 - 0.347862i$ \\
0.0034 & 4.587 & $0.432803 - 0.348633i$ \\
0.0034 & 4.597 & $0.433065 - 0.349404i$ \\
\bottomrule
\end{tabular}
\caption{$\alpha=0.1$}
\label{tab:4a}
\end{subtable}
\hfill
\begin{subtable}[t]{0.45\textwidth}
\centering
\begin{tabular}{ccc}
\toprule
$P$ & $r_h$ & $\omega$ \\
\midrule
0.0054 & 3.630 & $0.553016 - 0.435934i$ \\
0.0054 & 3.730 & $0.556998 - 0.448230i$ \\
0.0054 & 3.830 & $0.561110 - 0.460510i$ \\
0.0054 & 3.930 & $0.565345 - 0.472774i$ \\
0.0054 & 4.030 & $0.569700 - 0.485024i$ \\
\bottomrule
\end{tabular}
\caption{$\alpha=0.3$}
\label{tab:4b}
\end{subtable}
\vspace{1em}
\begin{subtable}[t]{0.45\textwidth}
\centering
\begin{tabular}{ccc}
\toprule
$P$ & $r_h$ & $\omega$ \\
\midrule
0.0089 & 2.832 & $0.725225 - 0.556628i$ \\
0.0089 & 2.932 & $0.731344 - 0.577002i$ \\
0.0089 & 3.032 & $0.737752 - 0.597339i$ \\
0.0089 & 3.132 & $0.744434 - 0.617642i$ \\
0.0089 & 3.232 & $0.751378 - 0.637914i$ \\
\bottomrule
\end{tabular}
\caption{$\alpha=0.5$}
\label{tab:4c}
\end{subtable}
\caption{The QNM frequencies of the large black hole branch as functions of $r_h$ at fixed pressure. (a) $\alpha=0.1$, (b) $\alpha=0.3$, (c) $\alpha=0.5$.}
\label{table4}
\end{table}

\begin{figure}[htbp]
\centering
\begin{subfigure}[t]{0.49\textwidth}
\centering
\includegraphics[width=\textwidth]{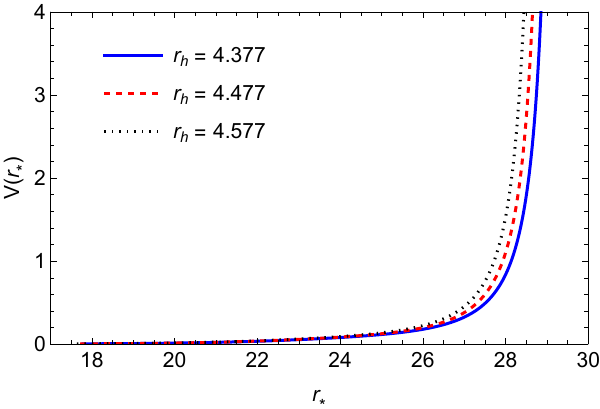}
\caption{$\alpha=0.1$}
\label{fig:5a}
\end{subfigure}
\hfill
\begin{subfigure}[t]{0.49\textwidth}
\centering
\includegraphics[width=\textwidth]{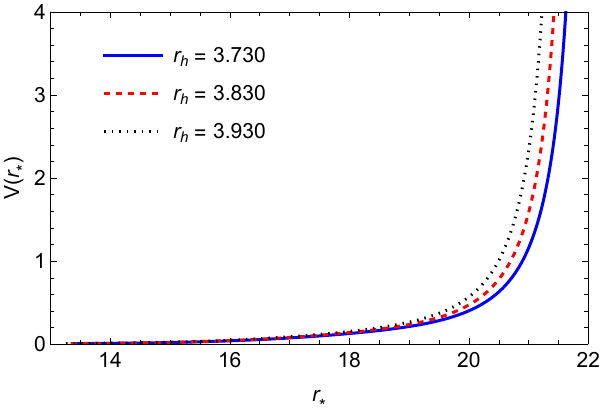}
\caption{$\alpha=0.3$}
\label{fig:5b}
\end{subfigure}
\vspace{1em}
\begin{subfigure}[t]{0.49\textwidth}
\centering
\includegraphics[width=\textwidth]{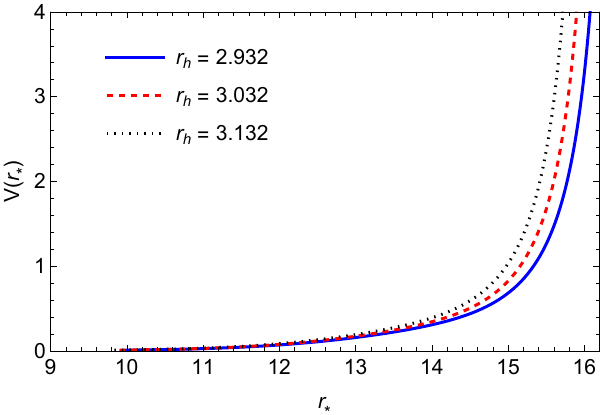}
\caption{$\alpha=0.5$}
\label{fig:5c}
\end{subfigure}
\caption{The effective potential distributions at fixed pressure for different values of the horizon radius $r_h$. (a) $\alpha=0.1$ with $P=0.0034$, (b) $\alpha=0.3$ with $P=0.0054$, (c) $\alpha=0.5$ with $P=0.0089$.}
\label{fig:5}
\end{figure}

Tables~\ref{table3} and \ref{table4} isolate the effect of the horizon radius by keeping the pressure, and hence the AdS radius $L$, fixed within each parameter set. On the small black hole branch, increasing $r_h$ leads to a slight decrease in $\operatorname{Re}(\omega)$ and an increase in $|\operatorname{Im}(\omega)|$. On the large black hole branch, both $\operatorname{Re}(\omega)$ and $|\operatorname{Im}(\omega)|$ increase with $r_h$, indicating a considerably stronger dependence on the horizon radius. This difference is also reflected in the effective potentials shown in Fig.~\ref{fig:5}: at fixed pressure, varying $r_h$ changes the shape and radial distribution of the potential, consistent with the more pronounced QNM response of the large black hole branch.

The comparison among different values of $\alpha$ shows that the PFDM background changes the overall scale of the QNM frequencies. For both branches, larger $\alpha$ generally corresponds to larger values of the real part and the magnitude of the imaginary part. Thus, at fixed pressure, a stronger PFDM background shifts the QNM spectrum toward larger $\operatorname{Re}(\omega)$ and larger $|\operatorname{Im}(\omega)|$, while the qualitative dependence on $r_h$ remains unchanged.

\begin{table}[htbp]
\centering
\begin{subtable}[t]{0.45\textwidth}
\centering
\begin{tabular}{ccc}
\toprule
$r_h$ & $P$ & $\omega$ \\
\midrule
0.9570 & 0.0038 & $0.417906 - 0.082369i$ \\
0.9570 & 0.0039 & $0.422497 - 0.085164i$ \\
0.9570 & 0.0040 & $0.427024 - 0.087965i$ \\
0.9570 & 0.0041 & $0.431491 - 0.090771i$ \\
0.9570 & 0.0042 & $0.435899 - 0.093582i$ \\
\bottomrule
\end{tabular}
\caption{$\alpha=0.1$}
\label{tab:5a}
\end{subtable}
\hfill
\begin{subtable}[t]{0.45\textwidth}
\centering
\begin{tabular}{ccc}
\toprule
$r_h$ & $P$ & $\omega$ \\
\midrule
0.8400 & 0.0057 & $0.530353 - 0.104271i$ \\
0.8400 & 0.0058 & $0.534324 - 0.106641i$ \\
0.8400 & 0.0059 & $0.538258 - 0.109014i$ \\
0.8400 & 0.0060 & $0.542156 - 0.111391i$ \\
0.8400 & 0.0061 & $0.546021 - 0.113772i$ \\
\bottomrule
\end{tabular}
\caption{$\alpha=0.3$}
\label{tab:5b}
\end{subtable}
\vspace{1em}
\begin{subtable}[t]{0.45\textwidth}
\centering
\begin{tabular}{ccc}
\toprule
$r_h$ & $P$ & $\omega$ \\
\midrule
0.7350 & 0.0091 & $0.697207 - 0.142711i$ \\
0.7350 & 0.0092 & $0.700555 - 0.144734i$ \\
0.7350 & 0.0093 & $0.703884 - 0.146676i$ \\
0.7350 & 0.0094 & $0.707194 - 0.148787i$ \\
0.7350 & 0.0095 & $0.710487 - 0.150817i$ \\
\bottomrule
\end{tabular}
\caption{$\alpha=0.5$}
\label{tab:5c}
\end{subtable}
\caption{The QNM frequencies of the small black hole branch as functions of $P$ at fixed horizon radius $r_h$. (a) $\alpha=0.1$, (b) $\alpha=0.3$, (c) $\alpha=0.5$.}
\label{table5}
\end{table}

\begin{table}[htbp]
\centering
\begin{subtable}[t]{0.45\textwidth}
\centering
\begin{tabular}{ccc}
\toprule
$r_h$ & $P$ & $\omega$ \\
\midrule
4.578 & 0.0032 & $0.416340 - 0.327227i$ \\
4.578 & 0.0033 & $0.424474 - 0.337585i$ \\
4.578 & 0.0034 & $0.432568 - 0.347939i$ \\
4.578 & 0.0035 & $0.440625 - 0.358289i$ \\
4.578 & 0.0036 & $0.448647 - 0.368635i$ \\
\bottomrule
\end{tabular}
\caption{$\alpha=0.1$}
\label{tab:6a}
\end{subtable}
\hfill
\begin{subtable}[t]{0.45\textwidth}
\centering
\begin{tabular}{ccc}
\toprule
$r_h$ & $P$ & $\omega$ \\
\midrule
3.830 & 0.0052 & $0.547670 - 0.443299i$ \\
3.830 & 0.0053 & $0.554399 - 0.451870i$ \\
3.830 & 0.0054 & $0.561110 - 0.460510i$ \\
3.830 & 0.0055 & $0.567801 - 0.469148i$ \\
3.830 & 0.0056 & $0.574475 - 0.477784i$ \\
\bottomrule
\end{tabular}
\caption{$\alpha=0.3$}
\label{tab:6b}
\end{subtable}
\vspace{1em}
\begin{subtable}[t]{0.45\textwidth}
\centering
\begin{tabular}{ccc}
\toprule
$r_h$ & $P$ & $\omega$ \\
\midrule
3.032 & 0.0087 & $0.727062 - 0.583682i$ \\
3.032 & 0.0088 & $0.732411 - 0.590511i$ \\
3.032 & 0.0089 & $0.737752 - 0.597339i$ \\
3.032 & 0.0090 & $0.743083 - 0.604167i$ \\
3.032 & 0.0091 & $0.748405 - 0.610994i$ \\
\bottomrule
\end{tabular}
\caption{$\alpha=0.5$}
\label{tab:6c}
\end{subtable}
\caption{The QNM frequencies of the large black hole branch as functions of $P$ at fixed horizon radius $r_h$. (a) $\alpha=0.1$, (b) $\alpha=0.3$, (c) $\alpha=0.5$.}
\label{table6}
\end{table}

Then, we isolate the effect of the pressure by fixing $r_h$ and varying $P$, as shown in Tables~\ref{table5} and \ref{table6}. On both the small and large black hole branches, increasing $P$ increases both $\operatorname{Re}(\omega)$ and $|\operatorname{Im}(\omega)|$. Since $P=3/(8\pi L^2)$, increasing the AdS radius $L$ instead decreases both the oscillation frequency and the damping rate.

The effects of $r_h$ and $P$ on the QNM spectrum are therefore different. At fixed pressure, increasing $r_h$ increases $|\operatorname{Im}(\omega)|$ on both branches, while $\operatorname{Re}(\omega)$ decreases slightly for small black holes but increases noticeably for large black holes. By contrast, at fixed $r_h$, increasing $P$ shifts both components of the QNM frequency upward on both branches.

\begin{figure}[htbp]
\centering
\begin{subfigure}[t]{0.48\textwidth}
\centering
\includegraphics[width=\textwidth]{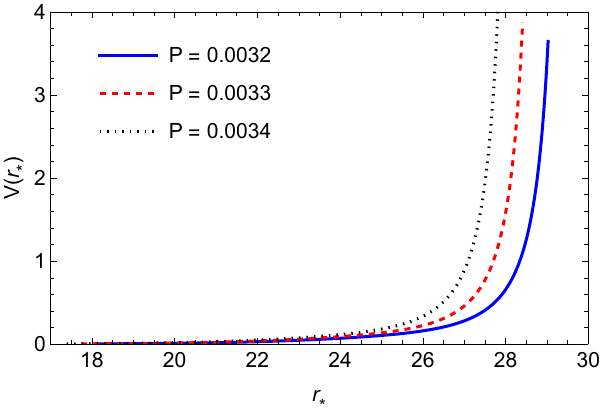}
\caption{$\alpha=0.1$}
\label{fig:6a}
\end{subfigure}
\hfill
\begin{subfigure}[t]{0.48\textwidth}
\centering
\includegraphics[width=\textwidth]{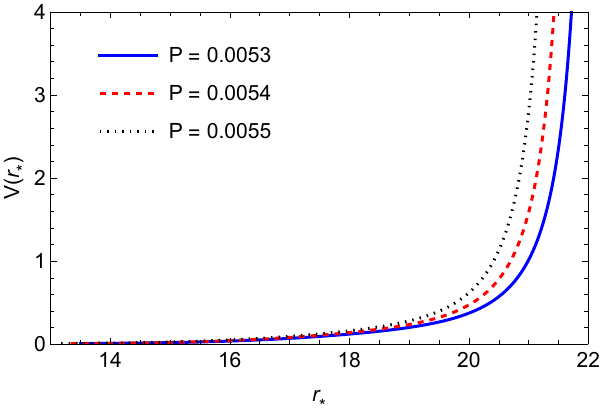}
\caption{$\alpha=0.3$}
\label{fig:6b}
\end{subfigure}
\vspace{1em}
\begin{subfigure}[t]{0.48\textwidth}
\centering
\includegraphics[width=\textwidth]{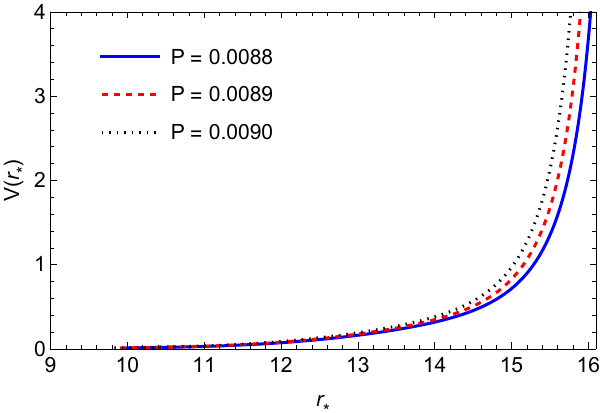}
\caption{$\alpha=0.5$}
\label{fig:6c}
\end{subfigure}
\caption{The effective potential distributions used to illustrate the effect of varying $P$, or equivalently the AdS radius $L$, at fixed horizon radius. (a) $\alpha=0.1$ with $r_h=4.578$, (b) $\alpha=0.3$ with $r_h=3.830$, (c) $\alpha=0.5$ with $r_h=3.032$.}
\label{fig:6}
\end{figure}

Figure~\ref{fig:6} qualitatively illustrates the pressure dependence for representative large black hole configurations. At fixed $r_h$, varying $P$, or equivalently $L$, changes the shape and radial scale of the effective potential. Increasing $L$ lowers the characteristic frequency scale of the perturbation, consistent with the decreases in both $\operatorname{Re}(\omega)$ and $|\operatorname{Im}(\omega)|$ shown in Table~\ref{table6}. The same numerical trend is found for the small black hole branch in Table~\ref{table5}.

The comparison among different values of $\alpha$ further shows that the PFDM background changes the overall scale of the QNM frequencies. For the representative points chosen near the transition region, larger $\alpha$ corresponds to larger pressure and smaller fixed horizon radius. Meanwhile, both $\operatorname{Re}(\omega)$ and $|\operatorname{Im}(\omega)|$ increase with $\alpha$. Thus, although the qualitative dependence on $P$ remains the same for different PFDM parameters, the PFDM background shifts the QNM spectrum toward larger $\operatorname{Re}(\omega)$ and larger $|\operatorname{Im}(\omega)|$.

To quantify the competition between the effects of $r_h$ and $P$, we expand the QNM frequency to leading order around a reference point $(r_h,P)$,
\begin{align}
\label{eq3.8}
\omega \left( {{r_h} + \Delta {r_h},P + \Delta P} \right) = \omega \left({r_h},P\right) + \frac{{\partial \omega }}{{\partial {r_h}}}\Delta {r_h} + \frac{{\partial \omega }}{{\partial P}}\Delta P + {\mathcal{O}}\left( {\Delta r_h^2,\Delta {P^2},\Delta {r_h}\Delta P} \right),
\end{align}
Here the two linear terms describe the contributions from the variation of the horizon radius and the pressure, respectively. For convenience, we define  ${\Delta _1} = \left( {{{\partial \omega } \mathord{\left/ {\vphantom {{\partial \omega } {\partial {r_h}}}} \right. \kern-\nulldelimiterspace} {\partial {r_h}}}} \right)\Delta {r_h}$ and ${\Delta _2} = \left( {{{\partial \omega } \mathord{\left/
 {\vphantom {{\partial \omega } {\partial P}}} \right.
 \kern-\nulldelimiterspace} {\partial P}}} \right)\Delta P$.

Along an isothermal path, the variations of $P$ and $r_h$ are constrained by the equation of state~(\ref{eq2.8}). Taking its differential at fixed temperature gives
\begin{equation}
\label{eq3.9}
{\rm d}P = \left( \frac{1}{4\pi r_h^3} - \frac{Q^2}{2\pi r_h^5} - \frac{T}{2r_h^2} + \frac{3\alpha}{8\pi r_h^4} \right) {\rm d}r_h.
\end{equation}
For a given pressure step $\Delta P$, the corresponding change in the  radius of horizon $\Delta r_h$, is estimated from Eq.~(\ref{eq3.9}). We then choose representative reference points near the transition region for each value of $\alpha$. The local derivatives with respect to $r_h$ and $P$ are estimated from Tables~\ref{table3}-\ref{table6}, and the resulting contributions, $\Delta_1=\partial_{r_h}\omega\,\Delta r_h$ and $\Delta_2=\partial_P\omega\,\Delta P$ are listed in Tables~\ref{table7}-\ref{table9}.

\begin{table}[htbp]
\centering
\begin{tabular}{ccccc}
\toprule
$r_h$ & $P$ & $\omega$ & $\Delta_1$ & $\Delta_2$ \\
\midrule
0.95700 & 0.00400 & $0.427024 - 0.087965i$ & $0$ & $0$ \\
0.96003 & 0.00390 & $0.422482 - 0.085303i$ & $-0.000014 - 0.000142i$ & $-0.004498 + 0.002803i$ \\
0.96306 & 0.00380 & $0.417876 - 0.082642i$ & $-0.000028 - 0.000283i$ & $-0.008997 + 0.005606i$ \\
0.96609 & 0.00370 & $0.413201 - 0.079983i$ & $-0.000042 - 0.000425i$ & $-0.013495 + 0.008410i$ \\
\midrule
4.43183 & 0.00345 & $0.432772 - 0.341676i$ & $-0.003807 + 0.011257i$ & $0.004038 - 0.005176i$ \\
4.57783 & 0.00340 & $0.432564 - 0.347926i$ & $0$ & $0$ \\
4.72383 & 0.00335 & $0.432304 - 0.353838i$ & $0.003807 - 0.011257i$ & $-0.004038 + 0.005176i$ \\
\bottomrule
\end{tabular}
\caption{The contributions of $r_h$ and $P$ to the QNM frequencies for $\alpha=0.1$. $\Delta_1$ and $\Delta_2$ denote the contributions from changes in $r_h$ and $P$, respectively, and $\omega$ is the linearly approximated QNM frequency. The upper and lower tables correspond to the small and large black hole branches, respectively.}
\label{table7}
\end{table}

\begin{table}[htbp]
\centering
\begin{tabular}{ccccc}
\toprule
$r_h$ & $P$ & $\omega$ & $\Delta_1$ & $\Delta_2$ \\
\midrule
0.84000  & 0.00590 & $0.538258 - 0.109014i$ & $0$ & $0$ \\
0.84190  & 0.00580 & $0.534302 - 0.106776i$ & $-0.000019 - 0.000137i$ & $-0.003917 + 0.002376i$ \\
0.84380  & 0.00570 & $0.530314 - 0.104539i$ & $-0.000038 - 0.000273i$ & $-0.007834 + 0.004752i$ \\
0.84570  & 0.00560 & $0.526285 - 0.102302i$ & $-0.000057 - 0.000410i$ & $-0.011751 + 0.007128i$ \\
\midrule
3.75560 & 0.00545 & $0.561342 - 0.455613i$ & $-0.003100 + 0.010990i$ & $0.003351 - 0.004320i$ \\
3.83000 & 0.00540 & $0.561110 - 0.460510i$ & $0$ & $0$ \\
3.90440 & 0.00535 & $0.560851 - 0.465235i$ & $0.003100 - 0.010990i$ & $-0.003351 + 0.004320i$ \\
\bottomrule
\end{tabular}
\caption{The contributions of $r_h$ and $P$ to the QNM frequencies for $\alpha=0.3$. $\Delta_1$ and $\Delta_2$ denote the contributions from changes in $r_h$ and $P$, respectively, and $\omega$ is the linearly approximated QNM frequency. The upper and lower tables correspond to the small and large black hole branches, respectively.}
\label{table8}
\end{table}

\begin{table}[htbp]
\centering
\begin{tabular}{ccccc}
\toprule
$r_h$ & $P$ & $\omega$ & $\Delta_1$ & $\Delta_2$ \\
\midrule
0.73500 & 0.00930 & $0.703884 - 0.146760i$ & $0$ & $0$ \\
0.73612 & 0.00920 & $0.700533 - 0.144862i$ & $-0.000022 - 0.000128i$ & $-0.003320 + 0.002027i$ \\
0.73724 & 0.00910 & $0.697163 - 0.142964i$ & $-0.000043 - 0.000257i$ & $-0.006640 + 0.004054i$ \\
0.73836 & 0.00900 & $0.693774 - 0.141067i$ & $-0.000065 - 0.000385i$ & $-0.009960 + 0.006081i$ \\
\midrule
2.99352 & 0.00895 & $0.737896 - 0.592889i$ & $-0.002516 + 0.007821i$ & $0.002668 - 0.003414i$ \\
3.03200 & 0.00890 & $0.737752 - 0.597339i$ & $0$ & $0$ \\
3.07048 & 0.00885 & $0.737599 - 0.601699i$ & $0.002516 - 0.007821i$ & $-0.002668 + 0.003414i$ \\
\bottomrule
\end{tabular}
\caption{The contributions of $r_h$ and $P$ to the QNM frequencies for $\alpha=0.5$. $\Delta_1$ and $\Delta_2$ denote the contributions from changes in $r_h$ and $P$, respectively, and $\omega$ is the linearly approximated QNM frequency. The upper and lower tables correspond to the small and large black hole branches, respectively.}
\label{table9}
\end{table}

Tables~\ref{table7}-\ref{table9} show the separate contributions of $r_h$ and $P$ for $\alpha=0.1$, $0.3$, and $0.5$, respectively. For each value of $\alpha$, we choose one representative point on the small black hole branch and one on the large black hole branch near the transition region. The pressure steps are taken as $\Delta P=-0.0001$ for the small black hole branch and $\Delta P=-0.00005$ for the large black hole branch. The corresponding $\Delta r_h$ is estimated from Eq.~(\ref{eq3.9}), and the local derivatives are obtained from Tables~\ref{table3}-\ref{table6}. The resulting first-order contributions $\Delta_1$ and $\Delta_2$ are then compared in Tables~\ref{table7}-\ref{table9}.

For the small black hole branch, the pressure contribution $\Delta_2$ is much larger than the horizon-radius contribution $\Delta_1$ for all three values of $\alpha$. This indicates that the monotonic decrease of both $\operatorname{Re}\left(\omega\right)$ and $\left|\operatorname{Im}\left(\omega\right)\right|$ along the isothermal path is mainly driven by the decrease in pressure, or equivalently by the increase in the AdS radius $L$. Thus, although $r_h$ also changes along the isotherm, the QNM evolution of the small black hole branch is dominated by the pressure effect.

For the large black hole branch, the two contributions are comparable and generally act in opposite directions. In the real part, $\Delta_1$ and $\Delta_2$ largely cancel each other, which explains why $\operatorname{Re}\left(\omega \right)$ changes only weakly along the isothermal path. In the imaginary part, the contribution from the increasing horizon radius is stronger, so the net damping rate increases even though the decreasing pressure tends to reduce it. The same qualitative mechanism appears for all three PFDM parameters considered here. Therefore, the isothermal QNM behavior is controlled by different mechanisms on the two branches. The small black hole branch is mainly pressure dominated, whereas the large black hole branch is governed by a competition between the horizon-radius and pressure effects.

\section{QNM behavior at the critical point}
\label{SECTIV}
After studying the QNM behavior below the critical point, we now turn to the critical point itself. Below the critical point, the QNM spectrum exhibits an abrupt change across the first-order small/large black hole phase transition. At the critical point, this first-order transition terminates and the two branches merge. It is therefore natural to ask whether the QNM spectrum still carries a distinct dynamical signal of the critical behavior. In this section, we examine the QNMs along the critical isobar $P=P_c$ and the critical isotherm $T=T_c$.

\subsection{QNM behavior on the critical isobar}
\begin{figure}[htbp]
\centering
\begin{subfigure}[t]{0.48\textwidth}
\centering
\includegraphics[width=\textwidth]{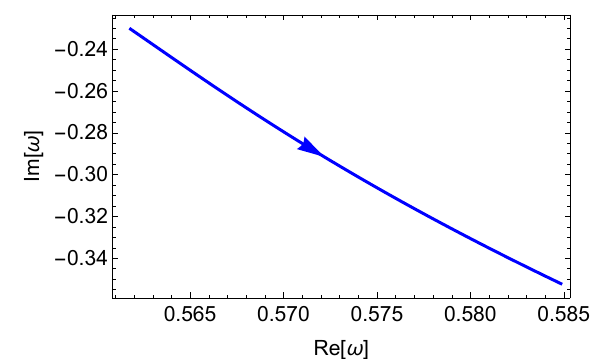}
\caption{$\alpha=0.1$}
\label{fig:7a}
\end{subfigure}
\hfill
\begin{subfigure}[t]{0.48\textwidth}
\centering
\includegraphics[width=\textwidth]{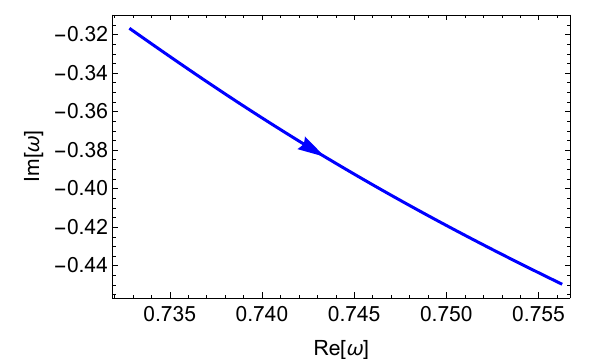}
\caption{$\alpha=0.3$}
\label{fig:7b}
\end{subfigure}
\vspace{1em}
\begin{subfigure}[t]{0.48\textwidth}
\centering
\includegraphics[width=\textwidth]{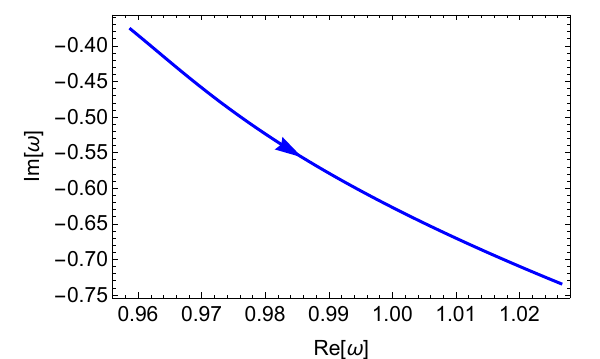}
\caption{$\alpha=0.5$}
\label{fig:7c}
\end{subfigure}
\caption{The QNM frequencies on the critical isobar $P=P_c$. The arrows indicate the direction of increasing horizon radius and temperature. No disconnected small/large black hole branches appear at the critical point. (a) $\alpha=0.1$, (b) $\alpha=0.3$, (c) $\alpha=0.5$.}
\label{fig:7}
\end{figure}

Along the critical isobar $P=P_c$, we compute the QNM frequencies for different values of the PFDM parameter, as listed in Table~\ref{table10}. As the temperature $T$ increases, the radius of horizon $r_h$ passes through the critical region. The corresponding QNM trajectories are shown in Fig.~\ref{fig:7}. Unlike the case below the critical pressure, the QNM trajectory does not split into two disconnected small/large black hole branches and does not show an abrupt jump. This behavior is consistent with the fact that the first-order phase transition disappears at the critical point, where the distinction between the small and large black hole phases is lost.

For all three values of $\alpha$, both $\operatorname{Re}\left(\omega\right)$ and $\left|\operatorname{Im}\left(\omega \right)\right|$ increase monotonically as $r_h$ increases along the critical isobar. The horizontal line in Table~\ref{table10} only separates the data below and above the critical radius $r_c$. It should not be interpreted as a boundary between two distinct thermodynamic phases. A comparison among different PFDM parameters shows that larger $\alpha$ shifts the QNM frequencies to larger values of both $\operatorname{Re}\left(\omega\right)$ and $\left|\operatorname{Im}\left(\omega\right)\right|$, corresponding to faster oscillation and stronger damping. However, this change only modifies the overall frequency scale and does not generate a discontinuous QNM signal at the critical point.

\begin{table}[htbp]
\centering
\begin{subtable}[t]{0.45\textwidth}
\centering
\begin{tabular}{ccc}
\toprule
$T$ & $r_h$ & $\omega$ \\
\midrule
0.0651 & 1.380758 & $0.562370 - 0.233866i$ \\
0.0652 & 1.410190 & $0.563071 - 0.238207i$ \\
0.0653 & 1.450940 & $0.564058 - 0.244301i$ \\
0.0654 & 1.623309 & $0.568487 - 0.270890i$ \\
\midrule
0.0655 & 1.913091 & $0.577198 - 0.317270i$ \\
0.0656 & 2.013684 & $0.580656 - 0.333620i$ \\
0.0657 & 2.080261 & $0.583073 - 0.344482i$ \\
0.0658 & 2.135721 & $0.585164 - 0.353548i$ \\
\bottomrule
\end{tabular}
\caption{$\alpha=0.1$}
\label{tab:10a}
\end{subtable}
\hfill
\begin{subtable}[t]{0.45\textwidth}
\centering
\begin{tabular}{ccc}
\toprule
$T$ & $r_h$ & $\omega$ \\
\midrule
0.0848 & 1.209005 & $0.732878 - 0.317264i$ \\
0.0849 & 1.234464 & $0.733767 - 0.323234i$ \\
0.0850 & 1.271848 & $0.735102 - 0.332119i$ \\
0.0851 & 1.373724 & $0.738959 - 0.356900i$ \\
\midrule
0.0852 & 1.639563 & $0.750927 - 0.423873i$ \\
0.0853 & 1.701057 & $0.754140 - 0.439627i$ \\
0.0854 & 1.745654 & $0.756579 - 0.451089i$ \\
0.0855 & 1.782202 & $0.758646 - 0.460500i$ \\
\bottomrule
\end{tabular}
\caption{$\alpha=0.3$}
\label{tab:10b}
\end{subtable}
\vspace{1em}
\begin{subtable}[t]{0.45\textwidth}
\centering
\begin{tabular}{ccc}
\toprule
$T$ & $r_h$ & $\omega$ \\
\midrule
0.1080 & 0.870322 & $0.957767 - 0.368934i$ \\
0.1090 & 0.895258 & $0.958838 - 0.376721i$ \\
0.1100 & 0.928121 & $0.960297 - 0.387445i$ \\
0.1110 & 0.977799 & $0.962583 - 0.404478i$ \\
\midrule
0.1120 & 1.275814 & $0.979044 - 0.518014i$ \\
0.1130 & 1.590434 & $1.004210 - 0.645783i$ \\
0.1140 & 1.716340 & $1.016800 - 0.697570i$ \\
0.1150 & 1.811694 & $1.027260 - 0.736886i$ \\
\bottomrule
\end{tabular}
\caption{$\alpha=0.5$}
\label{tab:10c}
\end{subtable}
\caption{The QNM frequencies on the critical isobar $P=P_c$. The horizontal line separates the data below and above the critical radius $r_c$ and does not indicate two distinct thermodynamic phases. (a) $\alpha=0.1$, (b) $\alpha=0.3$, (c) $\alpha=0.5$.}
\label{table10}
\end{table}

\subsection{QNM behavior on the critical isotherm}

\begin{figure}[htbp]
\centering
\begin{subfigure}[t]{0.48\textwidth}
\centering
\includegraphics[width=\textwidth]{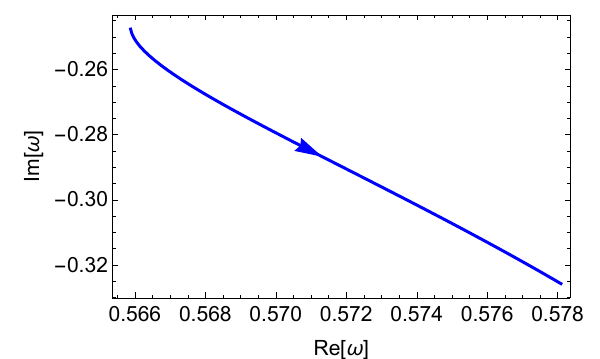}
\caption{$\alpha=0.1$}
\label{fig:8a}
\end{subfigure}
\hfill
\begin{subfigure}[t]{0.48\textwidth}
\centering
\includegraphics[width=\textwidth]{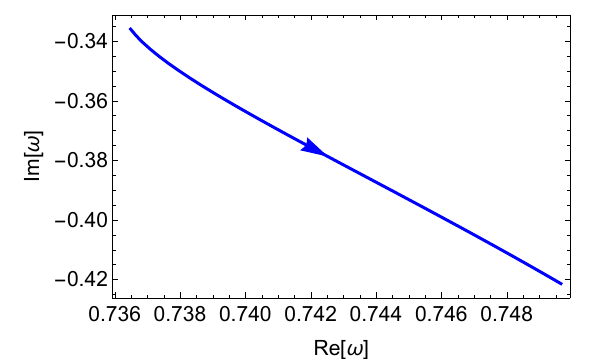}
\caption{$\alpha=0.3$}
\label{fig:8b}
\end{subfigure}
\vspace{1em}
\begin{subfigure}[t]{0.48\textwidth}
\centering
\includegraphics[width=\textwidth]{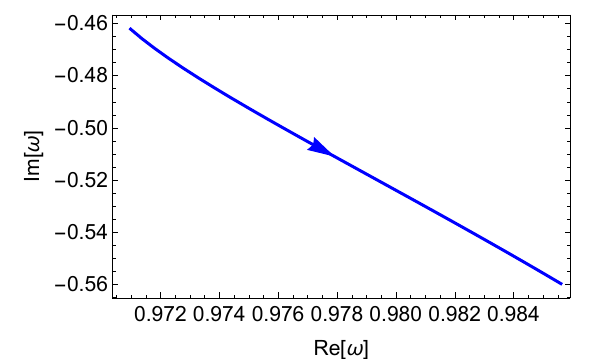}
\caption{$\alpha=0.5$}
\label{fig:8c}
\end{subfigure}
\caption{The QNM frequencies on the critical isotherm $T=T_c$. The arrows indicate the direction of increasing horizon radius and decreasing pressure. (a) $\alpha=0.1$, (b) $\alpha=0.3$, (c) $\alpha=0.5$.}
\label{fig:8}
\end{figure}

Along the critical isotherm $T=T_c$, we also compute the QNMs for different values of the PFDM parameter, as listed in Table~\ref{table11}. As the pressure $P$ decreases, the horizon radius $r_h$ increases, and the QNM frequencies vary smoothly with $r_h$, as shown in Fig.~\ref{fig:8}. Similar to the critical isobar, no separated small/large black hole branches or abrupt jump appears on the critical isotherm. This again indicates that the QNM spectrum does not show the discontinuous dynamical signal characteristic of the first-order phase transition at the critical point.

For all three values of $\alpha$, both $\operatorname{Re}(\omega)$ and $|\operatorname{Im}(\omega)|$ increase monotonically as $r_h$ increases along the critical isotherm. The horizontal line in Table~\ref{table11} only separates the data with $r_h<r_c$ from those with $r_h>r_c$, rather than two distinct thermodynamic phases. Comparing different PFDM parameters, a larger $\alpha$ shifts the QNM frequencies to larger values, indicating faster oscillation and stronger damping. This trend agrees with the behavior found on the critical isobar and in the first-order transition region, but it still does not generate any discontinuous QNM signal at the critical point.

\begin{table}[htbp]
\centering
\begin{subtable}[t]{0.45\textwidth}
\centering
\begin{tabular}{ccc}
\toprule
$P$ & $r_h$ & $\omega$ \\
\midrule
0.00739 & 1.463235 & $0.565878 - 0.247640i$ \\
0.00738 & 1.481015 & $0.565954 - 0.249987i$ \\
0.00737 & 1.503885 & $0.566160 - 0.253123i$ \\
0.00736 & 1.537320 & $0.566640 - 0.257905i$ \\
\midrule
0.00735 & 1.623309 & $0.568533 - 0.270936i$ \\
0.00734 & 1.876515 & $0.575638 - 0.310958i$ \\
0.00733 & 1.935713 & $0.577175 - 0.320084i$ \\
0.00732 & 1.978919 & $0.578227 - 0.326624i$ \\
\bottomrule
\end{tabular}
\caption{$\alpha=0.1$}
\label{tab:11a}
\end{subtable}
\hfill
\begin{subtable}[t]{0.45\textwidth}
\centering
\begin{tabular}{ccc}
\toprule
$P$ & $r_h$ & $\omega$ \\
\midrule
0.01160 & 1.284012 & $0.736487 - 0.335971i$ \\
0.01159 & 1.300486 & $0.736783 - 0.339638i$ \\
0.01158 & 1.323907 & $0.737348 - 0.345011i$ \\
0.01157 & 1.373685 & $0.738958 - 0.356890i$ \\
\midrule
0.01156 & 1.564520 & $0.746900 - 0.404380i$ \\
0.01155 & 1.607774 & $0.748645 - 0.415004i$ \\
0.01154 & 1.638221 & $0.749816 - 0.422383i$ \\
0.01153 & 1.662934 & $0.750727 - 0.428303i$ \\
\bottomrule
\end{tabular}
\caption{$\alpha=0.3$}
\label{tab:11b}
\end{subtable}
\vspace{1em}
\begin{subtable}[t]{0.45\textwidth}
\centering
\begin{tabular}{ccc}
\toprule
$P$ & $r_h$ & $\omega$ \\
\midrule
0.01837 & 1.132124 & $0.970987 - 0.462207i$ \\
0.01836 & 1.150155 & $0.971722 - 0.468896i$ \\
0.01835 & 1.191871 & $0.973859 - 0.484887i$ \\
0.01834 & 1.275814 & $0.979044 - 0.518014i$ \\
\midrule
0.01833 & 1.353654 & $0.983927 - 0.548702i$ \\
0.01832 & 1.374374 & $0.985158 - 0.556747i$ \\
0.01831 & 1.391186 & $0.986128 - 0.563221i$ \\
0.01830 & 1.405653 & $0.986938 - 0.568751i$ \\
\bottomrule
\end{tabular}
\caption{$\alpha=0.5$}
\label{tab:11c}
\end{subtable}
\caption{The QNM frequencies on the critical isotherm $T=T_c$. The horizontal line separates the data with $r_h<r_c$ from those with $r_h>r_c$. (a) $\alpha=0.1$, (b) $\alpha=0.3$, (c) $\alpha=0.5$.}
\label{table11}
\end{table}

In both the first-order phase transition region and at the critical point, $\alpha$ can affect the real part and the absolute value of the imaginary part of the QNM frequencies, making the perturbations oscillate faster and decay more rapidly. However, in the first-order transition region, $\alpha$ can enhance the abrupt slope change and the difference between the two phases in the QNM spectrum, thus amplifying the significance of the dynamical phase-transition signal. At the critical point, $\alpha$ cannot change the smooth and monotonic evolution of the QNMs, nor can it recover the dynamical signature of the phase transition; it can only influence the magnitude and rate of the evolution.

The origin of this difference lies in the fact that the impact of $\alpha$ on black hole thermodynamics is constrained by the critical conditions. In the first-order phase transition region, $\alpha$ can strengthen the dynamical signal by modulating the thermodynamic differences between small and large black holes. At the critical point, however, the thermodynamic distinction between the two phases vanishes, and the effect of $\alpha$ is restricted to the perturbation dynamics of a single phase, so it cannot recover the characteristic features of the phase transition. Even at the critical point, $\alpha$ can still influence the magnitude and pace of the QNM evolution, but it cannot restore the abrupt slope change or the two-phase branch that appears in the first-order transition. This demonstrates that, for the fundamental scalar mode and the paths studied here, the QNM spectrum is sensitive to the first-order phase transition but does not exhibit a comparably sharp signal at the second-order critical point. Whether other perturbation types or higher overtones could provide a more decisive signature remains an open question.

\section{QNM behavior along the coexistence curve}
\label{SECTV}
In Section~\ref{SECTIII}, we studied the isobaric and isothermal processes and showed that the QNM spectra exhibit abrupt changes when the system crosses the first-order small/large black hole phase transition point. In this section, we further examine the QNM behavior along the coexistence curve itself. The coexistence curve specifies the thermodynamic states where the small and large black hole phases have the same Gibbs free energy. Following this curve toward the critical point allows us to track how the dynamical difference between the two phases gradually disappears.

\begin{figure}[htbp]
\centering
\begin{subfigure}[t]{0.49\textwidth}
\centering
\includegraphics[width=\textwidth]{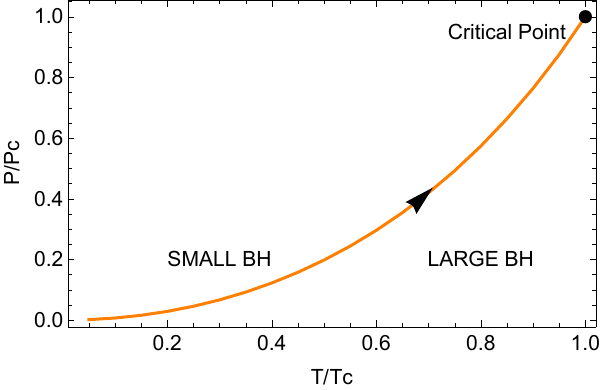}
\caption{$\alpha=0.1$}
\label{fig:9a}
\end{subfigure}
\hfill
\begin{subfigure}[t]{0.49\textwidth}
\centering
\includegraphics[width=\textwidth]{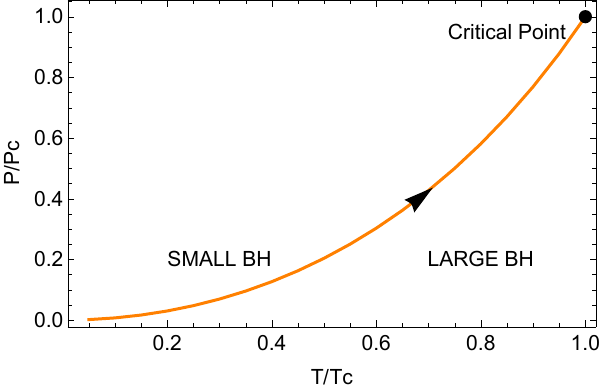}
\caption{$\alpha=0.3$}
\label{fig:9b}
\end{subfigure}
\vspace{1em}
\begin{subfigure}[t]{0.49\textwidth}
\centering
\includegraphics[width=\textwidth]{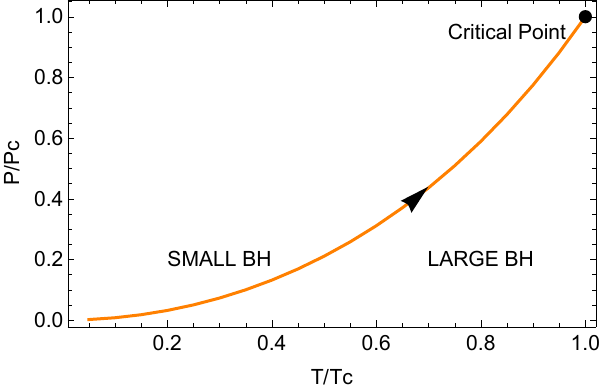}
\caption{$\alpha=0.5$}
\label{fig:9c}
\end{subfigure}
\caption{The coexistence curves of the small and large black hole phases. The arrows indicate the direction toward the critical point. (a) $\alpha=0.1$, (b) $\alpha=0.3$, (c) $\alpha=0.5$.}
\label{fig:9}
\end{figure}

\begin{figure}[htbp]
\centering
\begin{subfigure}[t]{0.47\textwidth}
\centering
\includegraphics[width=\textwidth]{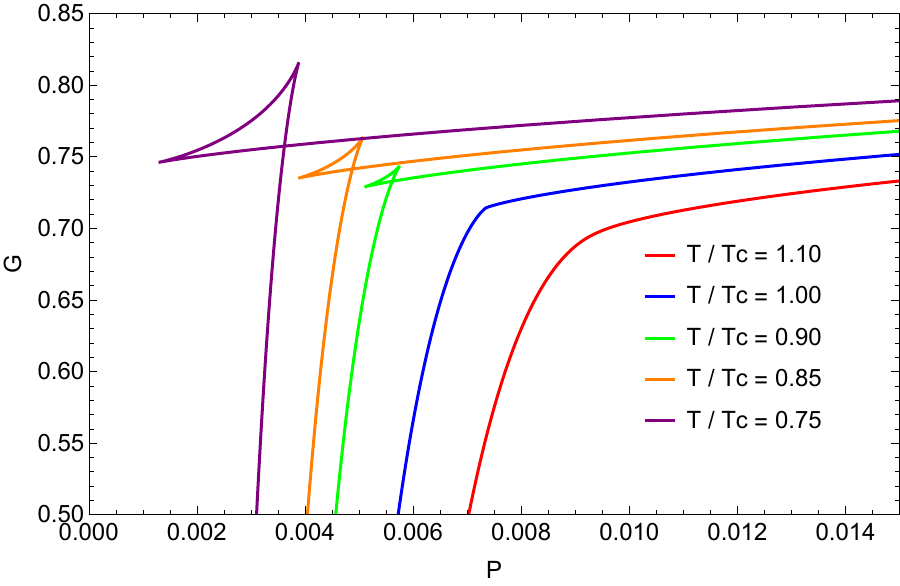}
\caption{$\alpha=0.1$}
\label{fig:10a}
\end{subfigure}
\hfill
\begin{subfigure}[t]{0.49\textwidth}
\centering
\includegraphics[width=\textwidth]{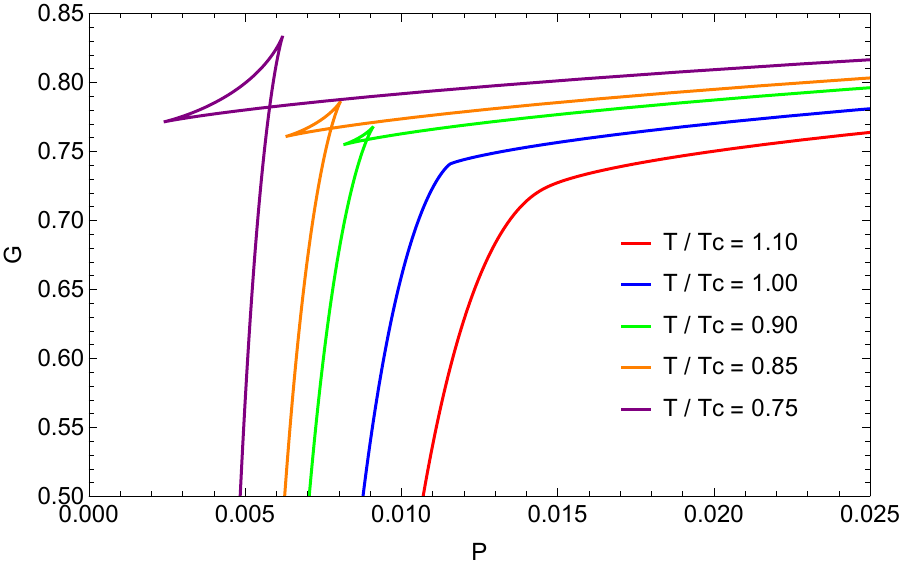}
\caption{$\alpha=0.3$}
\label{fig:10b}
\end{subfigure}
\vspace{1em}
\begin{subfigure}[t]{0.49\textwidth}
\centering
\includegraphics[width=\textwidth]{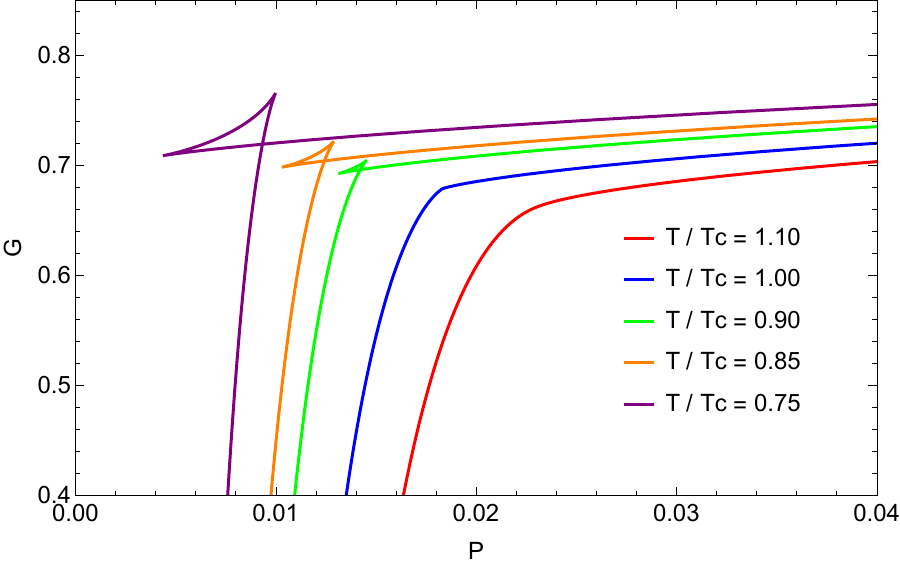}
\caption{$\alpha=0.5$}
\label{fig:10c}
\end{subfigure}
\caption{The swallowtail structure of the Gibbs free energy for the small/large black hole phase transition. (a) $\alpha=0.1$, (b) $\alpha=0.3$, (c) $\alpha=0.5$.}
\label{fig:10}
\end{figure}

The thermodynamic origin of the coexistence curve is reflected in the Gibbs free energy. As shown in Fig.~\ref{fig:10}, below the critical point the Gibbs free energy develops a swallowtail structure, which is the characteristic signal of a first-order small/large black hole phase transition. The crossing of the two stable branches corresponds to the coexistence of the two phases. As the critical point is approached, the swallowtail shrinks and finally disappears, indicating the termination of the first-order phase transition.

\begin{figure}[htbp]
\centering
\begin{subfigure}[t]{\textwidth}
\centering
\begin{tabular}{cc}
\includegraphics[width=0.4\textwidth]{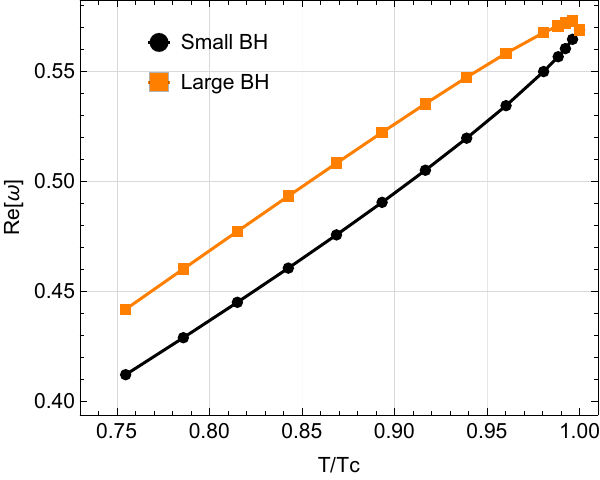} &
 \includegraphics[width=0.4\textwidth]{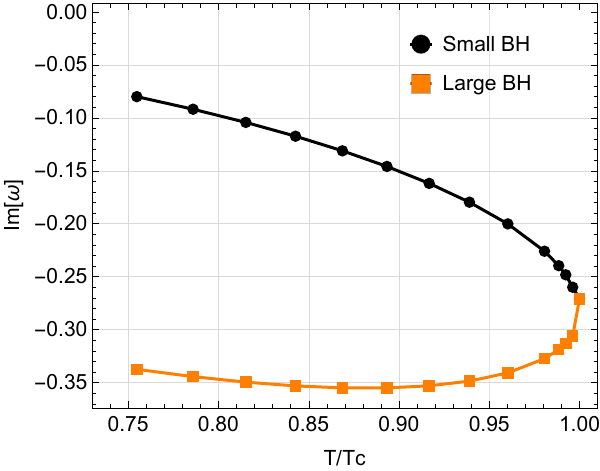}
\end{tabular}
\caption{$\alpha = 0.1$}
\label{fig:11a}
\end{subfigure}
\vspace{1em}
\begin{subfigure}[t]{\textwidth}
\centering
\begin{tabular}{cc}
\includegraphics[width=0.4\textwidth]{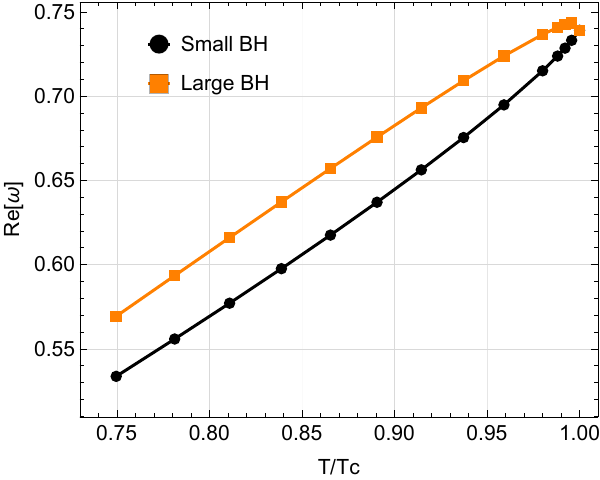} &
\includegraphics[width=0.4\textwidth]{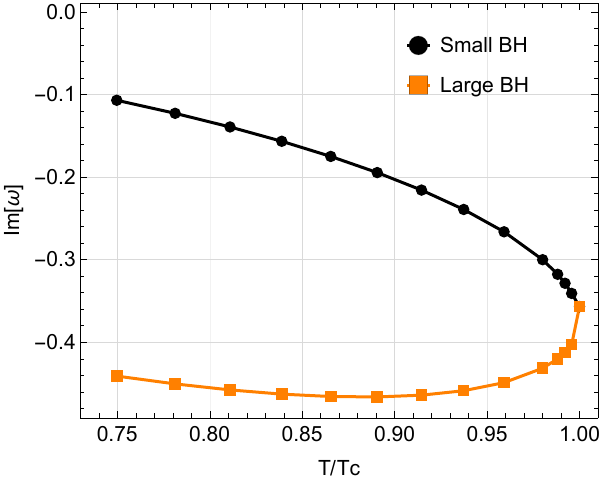}
\end{tabular}
\caption{$\alpha = 0.3$}
\label{fig:11b}
\end{subfigure}
\vspace{1em}
\begin{subfigure}[t]{\textwidth}
\centering
\begin{tabular}{cc}
\includegraphics[width=0.4\textwidth]{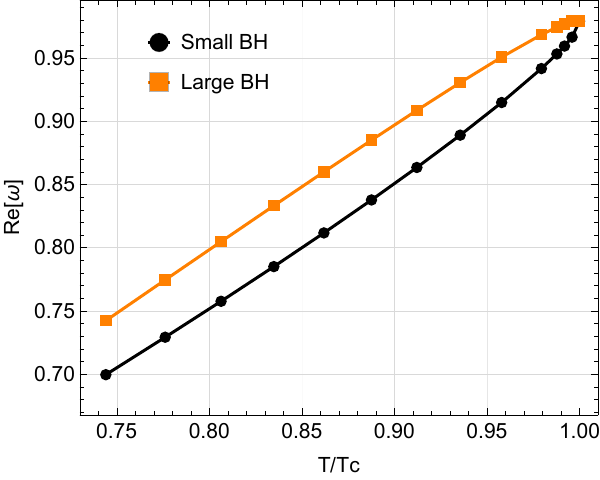} &
\includegraphics[width=0.4\textwidth]{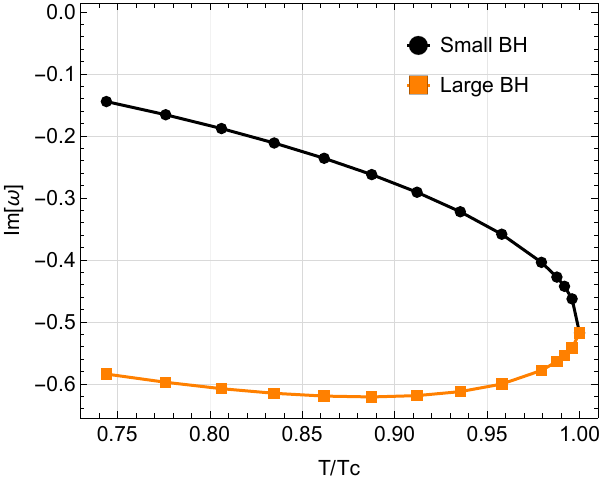}
\end{tabular}
\caption{$\alpha = 0.5$}
\label{fig:11c}
\end{subfigure}
\caption{The QNM frequencies along the coexistence curve. In each panel, the left plot shows $\operatorname{Re}(\omega)$ and the right plot shows $\operatorname{Im}(\omega)$ for the small and large black hole branches. (a) $\alpha=0.1$, (b) $\alpha=0.3$, (c) $\alpha=0.5$.}
\label{fig:11}
\end{figure}

Along the coexistence curve shown in Fig.~\ref{fig:9}, the corresponding QNM frequencies are displayed in Fig.~\ref{fig:11}. The imaginary part of the QNM frequency exhibits a clear separation between the small and large black hole branches over most of the coexistence curve, and this separation gradually shrinks as the critical point is approached. By contrast, the real parts of the two branches lie very close to each other along much of the coexistence curve and only begin to deviate noticeably near the critical point, before eventually merging there. This behavior is consistent with the thermodynamic picture shown in Fig.~\ref{fig:10}.

The PFDM parameter modifies this process quantitatively. For larger $\alpha$, the separation between the QNM spectra of the small and large black hole branches, especially in the imaginary part, remains more pronounced along the coexistence curve. Meanwhile, the thermodynamic phase structure is shifted to higher pressure. Thus, the PFDM background changes both the thermodynamic coexistence structure and the corresponding dynamical QNM response, while the qualitative tendency of the two QNM branches to approach each other near the critical point remains unchanged. This parallel behavior suggests a close correspondence between the thermodynamic and dynamical descriptions.

\section{Conclusion and Discussion}
\label{SECTVI}
In this work, we investigated the thermodynamic phase structure and QNM spectra of a charged AdS black hole surrounded by perfect fluid dark matter. Treating the cosmological constant as the thermodynamic pressure and the PFDM parameter $\alpha$ as an additional thermodynamic variable, we first reviewed the equation of state, the critical quantities, and the Gibbs free energy of the system. For the positive values of $\alpha$ considered here, the PFDM background shifts the critical point by increasing the critical temperature and pressure while decreasing the critical horizon radius. This shows that the surrounding dark matter environment changes the thermodynamic scale at which the small/large black hole phase transition occurs.

We then computed the fundamental massless scalar QNMs by the pseudospectral method and examined their behavior below the critical point. In both the isobaric and isothermal processes, the small and large black hole branches exhibit clearly separated QNM trajectories. When the system crosses the coexistence curve, the QNM spectrum changes from one branch to the other, and the two branches have distinct slopes in the complex frequency plane. This discontinuous behavior provides a dynamical signature of the first-order small/large black hole phase transition. The PFDM parameter changes the overall scale of the spectrum: larger $\alpha$ generally shifts the QNM frequencies toward larger $\operatorname{Re}(\omega)$ and larger $|\operatorname{Im}(\omega)|$.

For the isothermal process, we further separated the effects of the horizon radius and the pressure, or equivalently the AdS radius. At fixed pressure, the variation of $r_h$ mainly reflects the effect of the black hole size, while at fixed $r_h$ the variation of $P$ isolates the effect of the AdS scale. The first-order expansion in $\Delta r_h$ and $\Delta P$ shows that the QNM evolution of the small black hole branch is mainly controlled by the pressure contribution. For the large black hole branch, the two contributions compete more strongly: the imaginary part is more sensitive to the horizon-radius variation, whereas the real part is affected by a near cancellation between the $r_h$ and $P$ contributions. This explains why the QNM behavior in the isothermal process is not determined by the horizon radius alone.

At the critical point, the QNM spectra behave differently. Along both the critical isobar and the critical isotherm, the QNM frequencies vary smoothly and monotonically with the horizon radius. No abrupt jump or two-branch structure appears. This confirms that the QNM spectrum is sensitive to the first-order transition, but it does not provide an analogous sharp signal for the second-order critical point. Although increasing $\alpha$ still changes the magnitude of $\operatorname{Re}(\omega)$ and $|\operatorname{Im}(\omega)|$, it cannot restore the discontinuous dynamical feature that exists only below the critical point.

Finally, we studied the QNM evolution along the coexistence curve itself. The Gibbs free energy exhibits the usual swallowtail structure below the critical point, while the QNM spectra of the small and large black hole branches remain distinct along the coexistence curve. As the system approaches the critical point, the swallowtail structure shrinks and the difference between the two QNM branches gradually disappears. This parallel behavior suggests a close correspondence between the thermodynamic coexistence structure and the dynamical response encoded in QNMs. The PFDM parameter modifies this correspondence quantitatively by changing both the coexistence curve and the separation between the two QNM branches, while leaving the qualitative merging behavior near the critical point unchanged.

These results demonstrate that the fundamental scalar QNM spectrum is sensitive to the first-order small/large black hole phase transition, and this sensitivity persists in the presence of a PFDM environment. The QNM response can therefore be used as a dynamical supplement to the thermodynamic analysis. The analysis also shows that PFDM affects both the thermodynamic phase structure and the perturbative response of the black hole. Since QNMs are directly tied to black hole perturbation dynamics, their sensitivity to the surrounding matter distribution offers a useful theoretical tool for investigating how dark matter environments affect black hole phase transitions. Translating these results to gravitational perturbations, asymptotically flat geometries, and realistic dark matter distributions, and ultimately to observable gravitational‑wave signals, remains an open and intriguing challenge that deserves dedicated future investigation and discussion.

{\bf Acknowledgements}
We are very grateful to Yi Ling for helpful discussions.

\clearpage
\bibliography{Dynamic_TLN}
\bibliographystyle{JHEP}
\end{document}